\documentclass[fleqn,usenatbib,useAMS]{mnras}
\usepackage{fix-cm}

\usepackage{newtxtext,newtxmath}
\usepackage[T1]{fontenc}

\DeclareRobustCommand{\VAN}[3]{#2}
\let\VANthebibliography\thebibliography
\def\thebibliography{\DeclareRobustCommand{\VAN}[3]{##3}\VANthebibliography}

\usepackage{bm}
\usepackage{graphicx}	
\usepackage{amsmath}	

\usepackage{svg}
\usepackage{hhline}
\usepackage{ctable}
\usepackage{placeins}
\usepackage{siunitx}
\usepackage{arydshln}
\usepackage{threeparttable}

\DeclareRobustCommand{\linkedfootnotemark}[1]{%
  \hyperlink{Hfootnote.#1}{\footnotemark[#1]}%
}

\newcommand\secref[1]{Sect.~\ref{#1}}
\newcommand\figref[1]{Fig.~\ref{#1}}

\newcommand\equref[1]{Eq.~\eqref{#1}}
\newcommand\tabref[1]{Table~\ref{#1}}
\newcommand\appref[1]{Appendix~\ref{#1}}

\defcitealias{Sysoliatina2021}{SJ21}
\defcitealias{Cummings2018}{C18}
\usepackage{comment}

\title[Declining Milky Way SFH from CMD with WDs]{The full \textit{Gaia} colour--magnitude diagram with white dwarfs reveals a declining Milky Way star formation history}

\author[K. Sysoliatina and A. Vani]{
K. Sysoliatina$^{1,4}$\thanks{E-mail: ksysoliatina@aip.de, vani@mpa-garching.mpg.de}%
\thanks{Both authors contributed equally to this work.} and 
Akash Vani$^{2,3,4}$ \linkedfootnotemark{1}%
\linkedfootnotemark{2}
\\
$^{1}$ Leibniz Institute for Astrophysics Potsdam (AIP), An der Sternwarte 16, 14482 Potsdam, Germany\\
$^{2}$ Max Planck Institute for Astrophysics, Karl-Schwarzschild-Str. 1, 85741 Garching bei München, Germany   \\
$^{3}$ Ludwig-Maximilians-Universität München, Geschwister-Scholl-Platz 1, 80539 München, Germany \\
$^{4}$
   Astronomisches Rechen-Institut, Zentrum f\"{u}r Astronomie der Universit\"{a}t Heidelberg, 
M\"{o}nchhofstr. 12--14, 69120 Heidelberg, Germany \\
}

\date{Accepted XXX. Received YYY; in original form ZZZ}

\pubyear{\the\year{}}

\begin{document}
\label{firstpage}
\pagerange{\pageref{firstpage}--\pageref{lastpage}}
\maketitle

\begin{abstract}
White dwarfs (WDs) provide a fossil record of the Milky Way star formation history (SFH) through the combined effects of stellar evolution and cooling. They have, however, generally been modelled in isolation from the non-degenerate population, so that no single self-consistent set of Galactic assumptions has been required to reproduce both. We present \textsc{CALOMERA}, a semi-analytic chemo-dynamical population synthesis framework in which the vertical density profiles of mono-age subpopulations are computed self-consistently with the Galactic potential and coupled to modern WD cooling sequences, metallicity-dependent progenitor lifetimes and synthetic photometry. We calibrate 18 parameters describing the initial mass function, SFH, initial--final mass relation and WD population against a \textit{Gaia} DR3 sample of $7.98\times10^{6}$ stars within 1 kpc, including $1.1\times10^{5}$ WDs. The calibrated model reproduces the star counts in each colour--magnitude diagram region to within 3\% and the total number count to within 0.5\%, compared with the 81\% excess in the predicted WD count obtained with the previous main-sequence-only calibration. The recovered SFH peaks early, declines exponentially thereafter, and includes two recent bursts, located at $\tau\approx2.4$ and $0.5$\,Gyr. Three cooling libraries (Montr\'{e}al, BaSTI and LPCODE) yield comparably good fits, with differences confined to ${\rm M_G}\gtrsim15$~mag. The predicted luminosity function is in reasonable agreement with the 100 pc \textit{Gaia} WD sample. We further predict the Galactic WD distribution and differential number counts for surveys of varying depth. Extending the framework to the substellar regime using ATMO models, we predict the Galactic L, T, and Y dwarf populations and their contribution to the deep \textit{JWST} fields.
\end{abstract}




\section{Introduction}
White dwarfs (WDs) are the end products of the evolution of the vast majority of stars. Because all stars with initial masses below $\sim$8--11\,M$_\odot$ end their lives as WDs, the local WD population samples nearly every generation of star formation in the Milky Way (MW) disc and therefore constitutes a fundamental fossil record of its evolution. Their use as Galactic clocks rests on the fact that, once nuclear burning ceases, WD evolution is governed by cooling at essentially constant radius, a process that can be modelled from first principles \citep{Mestel1952}. Consequently, the faint end of the white dwarf luminosity function (WDLF) has long been used to estimate the age of the Galactic disc \citep[e.g.,][]{Schmidt1959, Liebert1988, Rowell2013, Tremblay2014}, while the detailed distribution of WDs in the colour--magnitude diagram (CMD) encodes information about the MW star formation history (SFH), stellar mass loss, and Galactic structure. Over the past two decades, rapid progress in WD observations, culminating in the era of the \textit{Gaia} mission \citep{Gaia2016}, together with major theoretical advances, has transformed this classical framework into a precision tool for Galactic archaeology \citep[e.g.,][]{Tremblay2024}.

Observationally, WD studies have progressed from small, extremely local spectroscopic samples confined to the nearest few tens of parsecs \citep[e.g.,][]{McCook1999, Gianninas2011, Limoges2013} to homogeneous wide-area surveys. The Sloan Digital Sky Survey \citep[SDSS,][]{York2000_SDSS} increased the known spectroscopic WD sample by an order of magnitude, enabling population-level studies of atmospheric types, mass distributions, and luminosity functions \citep[e.g.,][]{Kleinman2004, Kepler2015}. The decisive step came with \textit{Gaia}, whose astrometry and photometry resolved the WD sequence in the CMD with unprecedented clarity, expanded the census to more than 350\,000 candidates \citep{GentileFusillo2021}, and enabled nearly complete volume-limited samples in the Solar neighbourhood \citep[e.g.,][]{GentileFusillo2021, Golovin2023, OBrien2024}. These data also revealed rich structure within the WD sequence, including signatures of crystallisation, spectral evolution, unresolved binaries, and mass-dependent cooling \citep[e.g.,][]{Tremblay2019, Bedard2020, ElBadry2021}. Ongoing spectroscopic programmes with the Dark Energy Spectroscopic Instrument \citep[DESI,][]{Levi2019_DESI} and the WHT Enhanced Area Velocity Explorer \citep[WEAVE,][]{Jin2024_WEAVE} are extending these samples further. Deeper photometric and spectroscopic coverage is expected from the 4-metre Multi-Object Spectroscopic Telescope \citep[4MOST,][]{deJong2012_4MOST} and the Rubin Observatory Legacy Survey of Space and Time \citep[LSST,][]{Ivezic2019_LSST}, which together will turn WDs from a sparse, poorly accessible population into one of the principal tools of Galactic stellar population studies.

In parallel, WD evolutionary theory has advanced substantially. Modern cooling models treat neutrino losses at high luminosities, gravothermal cooling at intermediate stages, and the effects of core crystallisation, e.g., latent heat release, chemical phase separation, and the sedimentation of heavy species such as $^{22}$Ne, all of which introduce cooling delays at advanced ages \citep{Althaus2010, Salaris2010, Garcia_Berro2010, Tremblay2019, Camisassa2021}. Improved conductive opacities \citep{Blouin2020}, more accurate treatments of convective coupling in the envelope \citep{Tremblay2015}, element diffusion \citep{Bauer2019, Rolland2020}, and high-density atmospheric physics \citep{Blouin2019} have further refined cooling time-scales \citep[see also][]{Sahu2025}. Together with improved constraints on the initial--final mass relation \citep[IFMR; e.g.,][]{Weidemann2000, Cummings2018, Miller2026}, which links a progenitor to its WD mass, these developments have made WD ages and masses considerably more robust than they were two decades ago \citep[e.g.,][]{Fontaine2001, Althaus2010, Salaris2010, Tremblay2024}. As a result, WDs have become powerful and independent tracers of the age, SFH, and evolution of the Milky Way \citep[e.g.,][]{GarciaBerro2016, Fantin2019, Roberts2025}.

The MW star formation history is imprinted in its stellar populations, but directly inferring it from star counts is not straightforward because robust stellar age determination remains challenging. Indirect reconstruction techniques have therefore been developed that do not require
individual ages. Traditional approaches include fitting the CMD \citep[e.g.,][]{Mor2019, RuizLara2020, AlcazarJulia2025, AlzateTrujillo2026} and modelling chemical abundance distributions and enrichment laws \citep{RochaPinto2000, Snaith2015}. Although details differ between methods, many studies converge on a broadly declining SFH, with enhanced star formation at early times ($\sim$8--13 Gyr ago), a period of suppressed activity at $\sim$6--8 Gyr, and renewed star formation more recently, in some cases with multiple bursts \citep[e.g.,][]{Mor2019, RuizLara2020, Sysoliatina2021}.

WD-based reconstructions, by contrast, do not yield a unique picture of the local SFH \citep{Roberts2025}. \citet{Cukanovaite2023} found that an approximately constant formation rate over the past $\simeq10.5$~Gyr satisfactorily reproduces the 40~pc WD sample. \citet{Fantin2019}, however, inferred that disc star formation began $\sim11$~Gyr ago and reached a maximum of $\sim8.8\,\mathrm{M_\odot\,yr^{-1}}$ at a look-back time of $\sim9.8$~Gyr, followed by a gradual decline and a modest increase during the past $\simeq3$~Gyr. More structured solutions have also been recovered, e.g., \citet{Rowell2013} recovered two broad episodes of enhanced star formation at look-back times of $2$--$3$ and $7$--$9$~Gyr, separated by a period during which the inferred SFR declined by approximately $30$--$90$\%, while \citet{Isern2019} identified maxima at approximately $6$--$7$ and $2$--$3$~Gyr. Recent analyses likewise recover time-dependent solutions, although the detailed amplitudes and oldest age bins remain sensitive to the adopted assumptions \citep{Lam2026}. Thus, reconstructions from WD samples span histories ranging from nearly constant star formation to variations approaching an order of magnitude between adjacent epochs. This spread indicates that inferred WD-based SFHs remain sensitive to choices such as the cooling sequences, sample selection, atmospheric composition, IFMR, and treatment of Galactic structure, rather than being determined by the observations alone.

This motivates a closer examination of WD population-synthesis methodology. Synthetic stellar populations are generally constructed by combining an SFH, an initial mass function (IMF), an age--metallicity relation (AMR), and stellar evolutionary tracks with assumed spatial density laws for the thin disc, thick disc, and halo; the resulting densities are then projected onto observables such as the CMD or the luminosity function \citep{Robin2003_BGM, Bovy2016}. Several standard assumptions in this chain can introduce systematic bias. Exponential vertical density laws remain widely used despite the observed disc profile having a core near the Galactic plane \citep{Just2010, Sysoliatina2018}; this deviation matters most precisely in the small local volumes of 50--100 pc that are typically used to study WDs. Likewise, many population-synthesis studies adopt a single scale height for an entire Galactic component, neglecting the increase in scale height with stellar age caused by dynamical heating \citep{Just2010, Sysoliatina2018, Sysoliatina2021}. Because older populations occupy systematically larger vertical scale heights than younger populations, age-independent density laws bias the relative numbers of old and young stars in volume-limited samples and consequently affect the recovered SFH. Although these issues are recognised, relatively few studies model stellar populations within frameworks in which the density distributions are constrained self-consistently by the Galactic gravitational potential rather than imposed analytically \citep{Just2010, Sysoliatina2018, Robin2022_BGM}.

Further uncertainties are specific to WDs, whose total ages are reconstructed from a cooling model, an IFMR, and a progenitor lifetime. The IFMR encapsulates the cumulative and still poorly constrained effect of mass loss from the red giant branch through the thermally pulsing asymptotic giant branch (TP-AGB) phase, before the remnant settles onto the WD cooling sequence \citep[e.g.,][]{Weidemann2000, Cummings2018, Miller2026}. Modelled WD ages therefore inherit uncertainties from the IFMR, from the metallicity dependence of progenitor lifetimes, and from the adopted cooling sequences. Differences between these theoretical inputs, together with common simplifications such as neglecting the metallicity dependence of the main-sequence lifetime, contribute to a total systematic uncertainty that is rarely quantified. Moreover, WD populations are frequently modelled in isolation, without requiring the inferred SFH to reproduce the earlier evolutionary stages in the CMD under the same assumptions about the IMF, chemical evolution, and Galactic structure, which makes it difficult to distinguish genuine SFH signatures from WD-specific modelling artefacts \citep[e.g.,][]{Roberts2025}.

Existing Galactic population synthesis frameworks address these requirements to differing degrees. The Besan\c{c}on Galaxy Model \citep[BGM, ][]{Robin2003_BGM,Robin2022_BGM}, \texttt{TRILEGAL} \citep{Girardi2005_TRILEGAL}, and \texttt{GALAXIA} \citep{Sharma2011_Galaxia} generate synthetic catalogues spanning the full CMD, but do not simultaneously calibrate the white dwarf and non-degenerate populations using WD-specific model parameters. Conversely, dedicated WD population synthesis studies incorporate detailed cooling physics and explore the effects of the IFMR, SFH, and cooling models on the WD population \citep[e.g.,][]{Torres2016,Cukanovaite2023,Lam2026}, but generally adopt simplified prescriptions for the Galactic stellar distribution. To our knowledge, no existing framework combines a dynamically self-consistent Galactic model, age-dependent vertical structure, simultaneous modelling of degenerate and non-degenerate populations, metallicity-dependent stellar evolution, and Bayesian calibration of both Galactic and WD parameters within a single framework.

In this work, we aim to address these issues by modelling the WD population of the extended Solar neighbourhood within such a self-consistent Galactic framework. Our approach combines several elements that are generally not considered together: (i) realistic spatial density distributions calibrated within a dynamically self-consistent MW model; (ii) simultaneous modelling of WDs and non-degenerate stars (main sequence and giants) across the full CMD as observed by \textit{Gaia}; (iii) explicit treatment of the metallicity dependence of WD progenitor lifetimes; (iv) inclusion of WD model parameters in the fit, so that their uncertainties propagate into the recovered SFH; and (v) a comparison of three independent cooling libraries to quantify the systematic effect of this choice. Because the same calibrated framework maps stellar ages and masses onto observable properties for any evolutionary sequence, it can be naturally extended beyond white dwarfs to the substellar regime. Brown dwarfs probe a complementary mass range in which the Galactic SFH is inherited from the stellar population, while the predicted number counts are particularly sensitive to the low-mass IMF. As an additional application, we predict the Galactic population of L, T, and Y dwarfs, demonstrating that CALOMERA provides a unified population-synthesis framework spanning the main sequence, white dwarfs, and brown dwarfs. At the same time, the joint calibration of the stellar and WD populations reduces degeneracies in the inferred SFH, allowing us to separate genuine astrophysical features from modelling uncertainties more effectively than fits to either population alone.

This paper is structured as follows. \secref{Sec:Data} describes the observational data and sample selection. \secref{Sec:ModelDescription} presents our Galactic modelling framework and the ingredients for the synthesis of the WD population. \secref{Sec:ModelCalibration} outlines the details of the calibration of the model parameters. \secref{Sec:Results} presents the results for the reconstructed SFH and the model predictions for the WD population. \secref{Sec:Discussion} and \secref{Sec:Conclusions} present the discussion and conclusions, respectively.

\section{Data}
\label{Sec:Data}

For our analysis we use data from the \textit{Gaia} DR3 catalogue \citep{GaiaCollaboration2023_DR3}, from which we construct a local stellar sample using a combination of geometric and photometric criteria, described in detail below. In addition, we use three local white dwarf catalogues, the fifth catalogue of nearby stars \citep[CNS5;][]{Golovin2023}, the 40 pc WD sample of \citet{OBrien2024}, and the \textit{Gaia} catalogue of white dwarfs \citep[hereafter GCWD21;][]{GentileFusillo2021}, to test the calibrated model against samples constructed with different selection strategies (see \secref{Sec:Results}).

\subsection{Main sample}
\label{sec:main_sample}
\begin{figure*}
  \begin{minipage}[c]{0.8\textwidth}
    \includegraphics[width=\textwidth]{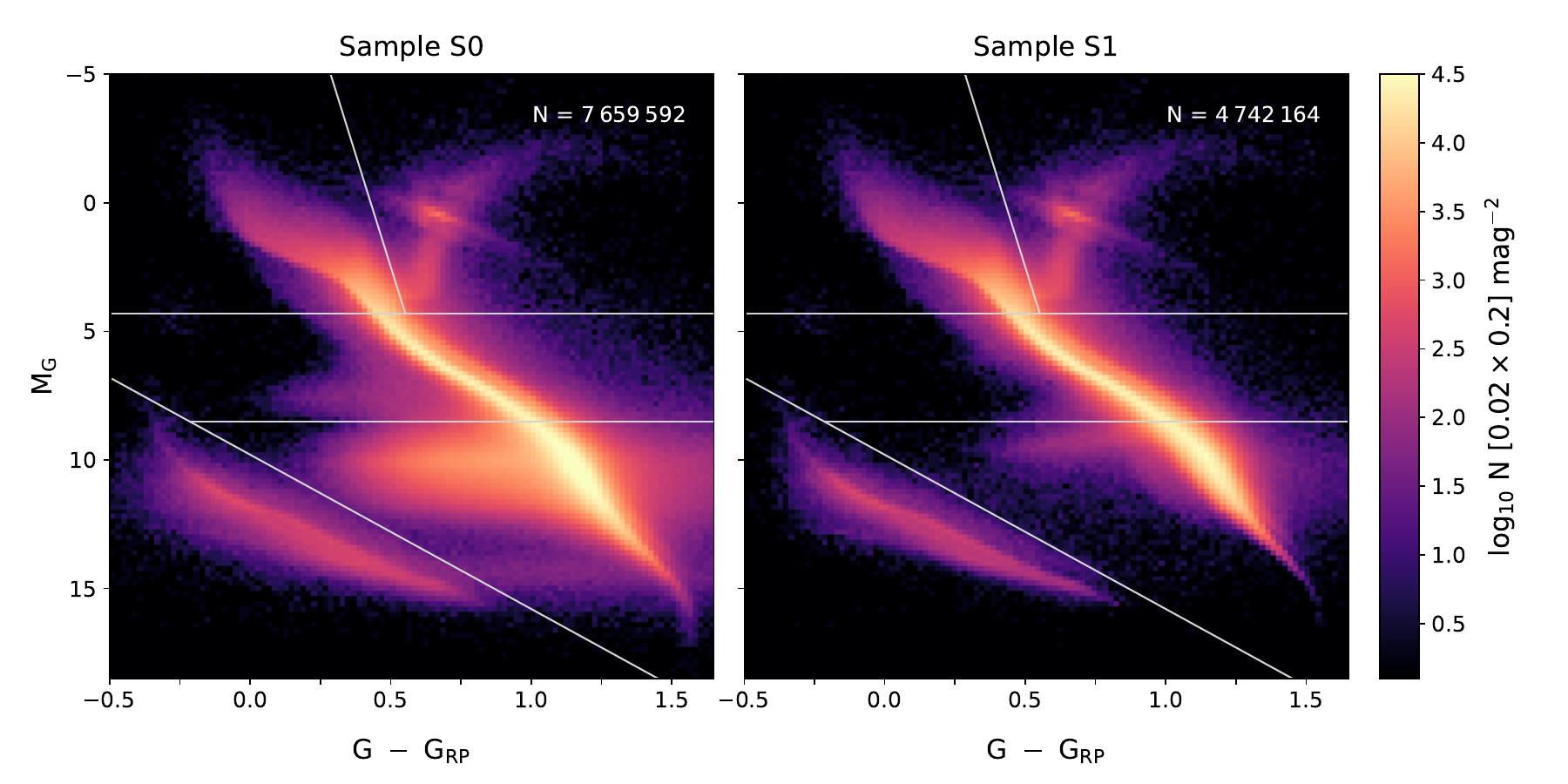}
  \end{minipage}\hfill
  \begin{minipage}[c]{0.18\textwidth}
    \caption{
        Hess diagrams of the selected samples. Grey lines show the boundaries of the CMD regions defined in \equref{eq:CMD_cuts}. \textit{Left:} full sample $S_0$. \textit{Right:} cleaned subsample $S_1$, with reduced binary and blending contamination.
    }
  \label{fig:data}
  \end{minipage}
\end{figure*}

\subsubsection{Geometric selection}

The geometric data selection follows \citet{Sysoliatina2021} (hereafter \citetalias{Sysoliatina2021}), ensuring spatial consistency with the previous calibration of our Galactic model. We restrict the sample to a narrow annulus around the Solar Galactocentric distance, adopting \mbox{$R_\odot = 8.275$ kpc} \citep{Abuter2021} and selecting stars within \mbox{$R_\odot \pm 150$ pc}. This minimises the effect of radial gradients in stellar density and ensures the validity of the local model assumptions. In the vertical direction we impose \mbox{$|z| < 150$ pc}, adopting a Solar height above the Galactic plane of \mbox{$z_\odot = 20$ pc} \citep{Bennett2019}. This restriction reduces the number of $z$-bins in the model and thus the computational cost of the MCMC analysis (see \secref{Sec:ModelCalibration}). Finally, we adopt a heliocentric distance limit of \mbox{$d \lesssim 1$ kpc}, which provides a sufficiently large sample while averaging over local density inhomogeneities. The resulting selection volume is a sphere of radius 1 kpc, intersected with a narrow Galactic annulus and truncated in height above and below the plane.

Distances are computed as inverse parallaxes. Although this approach becomes biased for distant or low signal-to-noise sources \citep{Luri2018}, and probabilistic methods are preferred in such cases \citep{Weiler2025}. However, for a local, high-quality \textit{Gaia} sample considered here, it provides reliable distances. We therefore restrict the sample to sources with positive parallaxes. Likewise, no parallax zero-point correction \citep{Lindegren2021} is applied, as the systematic offset is negligible compared to the parallaxes of sources within 150~pc.

To ensure reliable astrometry, we apply a quality cut of \mbox{$\texttt{astrometric\_excess\_noise} < 1$ mas}. We adopt this criterion because it directly identifies sources with poor astrometric solutions while avoiding the colour- and magnitude-dependent behaviour of the RUWE statistic, particularly for faint blue sources \citep{Lindegren2021}. The removed sources predominantly form a diffuse cloud in the CMD at absolute magnitudes fainter than $M_\mathrm{G} \approx 18$, indicative of unreliable astrometric solutions. Their removal is therefore not expected to introduce significant incompleteness.

\subsubsection{Photometric selection and sample definition}

We further restrict the sample to apparent magnitudes \mbox{$2 < G < 20.5$ mag}, where \textit{Gaia} DR3 is expected to be essentially complete \citep{CantatGaudin2023}. The resulting full sample, $S_0$, contains 7\,981\,600 stars. 

The full sample $S_0$ includes a substantial fraction of unresolved binaries and blended sources that populate the region between the MS and the WD sequence. As we do not explicitly model unresolved binaries, we construct a cleaned subsample, $S_1$, by applying additional photometric quality cuts:

\begin{align}
& \mathtt{phot\_rp\_mean\_flux\_over\_error} > 20 \nonumber \\
& \mathtt{phot\_bp\_mean\_flux\_over\_error} > 20 \nonumber \\
& \mathtt{phot\_g\_mean\_flux\_over\_error} > 50 \nonumber
\end{align}

This subsample contains 4\,744\,001 stars and is used for the comparison of normalised colour and magnitude distributions, while the full sample $S_0$ is used for the comparison of total number counts.

Because these cuts are based on photometric signal-to-noise, they act preferentially on faint sources, they retain 98.9, 100.0 and 96.6 per cent of the giant, upper main-sequence and main-sequence populations, respectively, but only 42.0\% of the lower main-sequence and 52.6\% of the WD population (\tabref{tab:StarCountsS0S1}). Consequently, $S_1$ is not a strictly binary-free sample, but rather a subsample with reduced binary and blending contamination and a well-defined magnitude-dependent selection function. The same photometric uncertainty model and flux-over-error selection criteria are applied to the synthetic catalogue, ensuring that the observational selection function is forward-modelled consistently (see \secref{sec:stellar_and_wd_popsynth}).

In this study we adopt the colour index $G - G_\mathrm{RP}$ rather than $G_\mathrm{BP} - G_\mathrm{RP}$. Although the latter provides a broader wavelength baseline, the $G_\mathrm{BP}$ fluxes of faint sources, and particularly of white dwarfs at \mbox{$M_\mathrm{G} \approx 10$--15 mag}, have signal-to-noise ratios (SNR) 5--20 times lower than in the $G$ band. Consequently, $G - G_\mathrm{RP}$ yields more stable colours in the faint regime relevant for our modelling.

\subsubsection{Deblending correction}

The \textit{Gaia} colour $G - G_\mathrm{RP}$ is affected by blending, since photometric contamination from close neighbours enters the broad $G$ and the narrower $G_\mathrm{RP}$ passbands disproportionately. To correct for this we apply the deblending procedure of \citet{Golovin2023}. That correction was calibrated over a limited range of $G_\mathrm{BP} - G_\mathrm{RP}$ colour and photometric SNR, whereas our sample extends partially beyond these ranges. To avoid artefacts at the boundary between the corrected and uncorrected regimes, we introduce a smooth transition between them, controlled by the weighting function

\begin{equation}
\omega(x) =
\begin{cases}
0, & x \le x_\mathrm{low} \\
0.5 \left[1 + \sin{\left(\pi \dfrac{x - C}{d}\right)}\right],
    & x_\mathrm{low} < x \le x_\mathrm{up} \\
1, & x > x_\mathrm{up}
\end{cases}
\end{equation} 

Here $d = x_\mathrm{up} - x_\mathrm{low}$ is the width of the transition interval and $C = (x_\mathrm{low} + x_\mathrm{up})/2$ is its centre, so that $\omega(x_\mathrm{low}) = 0$ and $\omega(x_\mathrm{up}) = 1$. The total weight
is

\begin{equation}
    \omega_\mathrm{deb} = \omega_\mathrm{col} \cdot \omega_\mathrm{SNR},
\end{equation}

with adopted transition intervals $(G_\mathrm{BP} - G_\mathrm{RP}) \in (0, 0.3)$ mag for $\omega_\mathrm{col}$ and $\mathrm{SNR} \in (20, 40)$ for $\omega_\mathrm{SNR}$. As the SNR entering $\omega_\mathrm{SNR}$ we adopt the smaller of the $G_\mathrm{BP}$ and $G_\mathrm{RP}$ values, $\min(\mathrm{SNR_{BP}}, \mathrm{SNR_{RP}})$.

The corrected colour is then
\begin{equation}
    (G - G_\mathrm{RP})_{\mathrm{corr}} =
    (1 - \omega_\mathrm{deb})(G - G_\mathrm{RP})_{\mathrm{obs}}
    + \omega_\mathrm{deb} \, (G - G_\mathrm{RP})_{\mathrm{deb}},
\end{equation}
where $(G - G_\mathrm{RP})_{\mathrm{obs}}$ is the observed colour and $(G - G_\mathrm{RP})_{\mathrm{deb}}$ the deblended colour following \citet{Golovin2023}. This construction avoids discontinuities in the CMD at the boundary between the two photometric regimes near $(G - G_\mathrm{RP}) \approx 0$ mag.

Finally, we correct for interstellar extinction and reddening using the three-dimensional dust map of \citet{Lallement2022}. The photometrically corrected CMDs of $S_0$ and $S_1$ are shown in \figref{fig:data}.

\begin{figure}
\centering
\includegraphics[width=1.0\linewidth]{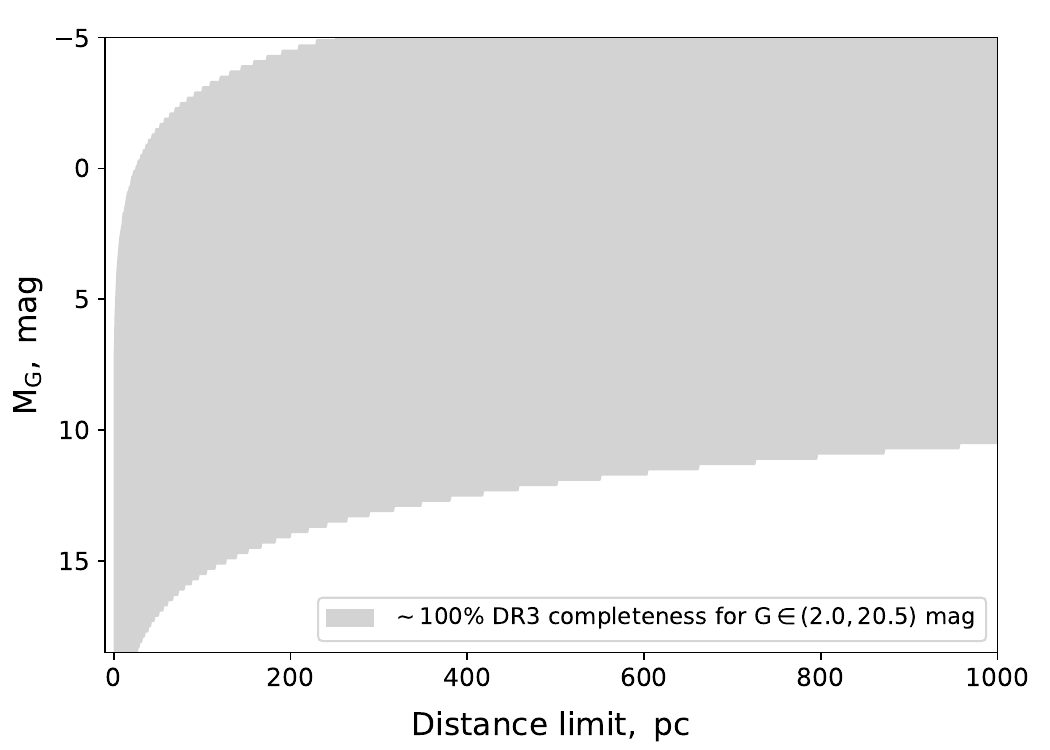}
\caption{Distance limits corresponding to the adopted apparent magnitude range
$2 < G < 20.5$ mag, as a function of absolute magnitude $M_\mathrm{G}$}
\label{fig:cmd_completeness}
\end{figure}

\subsubsection{Completeness}
\label{sec:completeness}

The adopted apparent magnitude limits translate into distance limits for a given absolute magnitude (\figref{fig:cmd_completeness}):
\begin{align}
    d_{\min}(M_\mathrm{G}) &= 10^{0.2\,(G_{\min} - M_\mathrm{G} + 5)} \ \mathrm{pc}, \\ 
    d_{\max}(M_\mathrm{G}) &= \min(10^{0.2\,(G_{\max} - M_\mathrm{G} + 5)}, 1000) \ \mathrm{pc}
    \label{eq:dmax}
\end{align}
Each absolute magnitude bin in the data is therefore sampled over a corresponding distance range, and in the modelling we reproduce this observational selection by assigning each synthetic population the corresponding volume. The maximum distance is truncated at 1~kpc to match our distance cut. A detailed description of how these limits are applied is given in \secref{sec:stellar_and_wd_popsynth}.

In computing the maximum observable distance, extinction is neglected. This approximation is justified because the sample is dominated by nearby sources, for which interstellar extinction is generally small. Furthermore, the synthetic photometry includes three-dimensional extinction when generating the mock catalogue (see \secref{sec:stellar_and_wd_popsynth}), such that extinction is accounted for in the forward modelling.

\subsubsection{CMD populations}

We divide the CMD into five regions corresponding to giants (G), upper main sequence (UMS), main sequence (MS), lower main sequence (LMS), and white dwarfs (WD), defined as:
\begin{subequations}
\label{eq:CMD_cuts}
\begin{align}
&\makebox[3em][l]{G:} &
 {}  {\rm M_G} \le {} \min(4.3, 35(G - G_\mathrm{RP}) - 15) \\
&\makebox[3em][l]{UMS:}&
 35\,(G - G_\mathrm{RP}) - 15 <  {\rm M_G}  < 4.3  \\
&\makebox[3em][l]{MS:} &
 4.3 < {\rm M_G}  < \min(8.5, 6(G - G_\mathrm{RP}) + 9.3) \\
&\makebox[3em][l]{LMS:} &
 8.5 \le {}  {\rm M_G}  < 6(G - G_\mathrm{RP}) + 9.3 \\
&\makebox[3em][l]{WD:} &
 {}  {\rm M_G} \ge 6(G - G_\mathrm{RP}) + 9.3
\end{align}
\end{subequations}

The number of stars in each region for $S_0$ and $S_1$ is given in \tabref{tab:StarCountsS0S1}. 
We note that the WD region is defined purely photometrically and will therefore contain some non-degenerate contaminants.

\begin{table}
    \centering
    \small
    \setlength{\tabcolsep}{4pt}
    \begin{tabular}{lllllll}
    \toprule
     & \multicolumn{1}{c}{G} & \multicolumn{1}{c}{UMS} & \multicolumn{1}{c}{MS} 
     & \multicolumn{1}{c}{LMS} & \multicolumn{1}{c}{WD} & \multicolumn{1}{c}{Total}\\
    \midrule
    $S_0$ & 84\,570 & 452\,741 & 2\,168\,620 & 4\,843\,148 & 110\,513 & 7\,659\,592 \\
    $S_1$ & 83\,723 & 452\,652 & 2\,095\,557 & 2\,051\,983 & 58\,249 & 4\,742\,164 \\
    \bottomrule
    \end{tabular}
    \normalsize
    \caption{Number of stars in each CMD region for the full sample $S_0$ and
    the cleaned subsample $S_1$.}
    \label{tab:StarCountsS0S1}
\end{table}

\subsection{Comparison with local volume-limited samples}
\label{sec:ValidationData}

To test the calibrated model against samples constructed with different selection strategies and completeness levels, we compare its predictions with three local WD catalogues, CNS5 \citep{Golovin2023}, containing 264 WDs within 25 pc; the 40 pc volume-limited sample of \citet{OBrien2024}, containing 1\,078 WDs; and GCWD21 \citep{GentileFusillo2021}. For GCWD21, we select sources with $P_{\rm WD}>0.75$ and measured parallaxes $\varpi\geq10$~mas, corresponding to nominal distances within 100~pc. This selection yields 16\,676 WD candidates. Approximately 82\% of sources in the inclusive $3\sigma_\varpi$ selection have relative parallax uncertainties below 10\%, supporting the use of a sharp parallax cut, although boundary scattering remains possible.

Although all three catalogues are based on \textit{Gaia} astrometry and photometry, they adopt different selection methodologies. CNS5 identifies nearby stars using a harmonic-amplitude criterion and provides a statistically complete census of WDs in the Solar neighbourhood \citep{Golovin2023}. GCWD21 identifies WD candidates probabilistically from \textit{Gaia} photometry and astrometry, assigning each source a WD probability, while the 40 pc catalogue of \citet{OBrien2024} is constructed from GCWD21 candidates supplemented by spectroscopic confirmation to produce a highly complete local WD sample.

The 25 pc and 40 pc catalogues are regarded as statistically volume complete for WDs and therefore provide robust constraints on the local WD number density. For the 100 pc GCWD21 sample, we estimate completeness using the $1/V_{\rm max}$ estimator when constructing its luminosity function.
When comparing the physical properties of WDs, we adopt the corrected effective temperatures, masses, and cooling ages of \citet{OBrien2024} for the 40 pc sample. For GCWD21 we consider both the hydrogen- and helium-atmosphere solutions reported in the catalogue. For the 25 pc sample we use the updated CNS6 compilation (Golovin et al., in preparation), in which the physical parameters are derived from multi-wavelength spectral energy distribution fitting.

Two caveats apply to these comparisons. First, all three catalogues are derived from the same underlying \textit{Gaia} data as our main sample, and all three volumes lie within our 1 kpc selection volume; they therefore provide a consistency check on the local normalisation and luminosity function rather than a fully independent validation. Second, the physical properties of WDs in all three samples were derived using the Montr\'{e}al WD model grids, so these comparisons cannot be used to discriminate between the cooling libraries considered in \secref{Sec:Results}.

\section{Model description} 
\label{Sec:ModelDescription}

\subsection{Model basics}
\label{sec:model_basic}

In this work we use a semi-analytic chemo-dynamical model of the Milky Way disc developed in a series of studies \citep{Just2010, Just2011, Rybizki2015, Sysoliatina2021, Sysoliatina2022}. We refer to this framework as \textsc{CALOMERA}\footnote{Chemo-dynAmicaL mOdel for Milky way Evolution and stellaR populAtion synthesis, named after the fast-moving tiger beetle genus \textit{Calomera}, reflecting the framework's emphasis on computational speed and efficiency.} (previously referred to as the JJ model), emphasising its role as a fast and flexible tool for Galactic population synthesis. \textsc{CALOMERA} retains the physical consistency of the previous developments while extending it to the joint modelling of white dwarf and non-degenerate stellar populations, extension to the substellar regime, together with a new MCMC calibration framework.

The model represents the Galactic disc as a superposition of isothermal mono-age stellar subpopulations with a time resolution of \mbox{25 Myr}. Each subpopulation is characterised by its age, metallicity, surface density, and vertical velocity dispersion. The stellar content and evolution are governed by four key input functions, namely, the SFH, the IMF, the age--velocity dispersion relation (AVR), and the age--metallicity relation (AMR).

A key feature of the model is its self-consistent treatment of the vertical MW structure. Instead of prescribing independent analytic density laws for different stellar populations, the vertical density profiles are derived consistently with the total Galactic gravitational potential. For an isothermal subpopulation of age $\tau_\mathrm{i}$ with vertical velocity dispersion $\sigma_W(\tau_\mathrm{i})$, assumed independent of $z$, the vertical density profile is\footnote{Throughout this paper, we use the following index notation: $i$ denotes mono-age isothermal subpopulations; $j$ denotes stellar assemblies, i.e., mono-(age, metallicity, mass) groups; $k$ labels the $z$-slices of the spatial grid; $l$ refers to the five CMD regions defined in \equref{eq:CMD_cuts}; $m$ labels the colour--absolute magnitude bins within each CMD region; and $s$ denotes the free model parameters.}
\begin{equation}
    \rho_\mathrm{i}(z) = \rho_\mathrm{i}(0)
    \exp\left[-\frac{\Phi(z)}{\sigma_W^2(\tau_\mathrm{i})}\right],
    \label{eq:rho_z}
\end{equation}
where $\rho_\mathrm{i}(0)$ is the in-plane density and $\Phi(z)$ the total vertical gravitational potential. This Boltzmann-like vertical law is a solution of the vertical Jeans equation with the tilt term neglected. The potential and density are obtained simultaneously by iteratively solving the one-dimensional Poisson equation,
\begin{equation}
    \frac{d^2\Phi(z)}{dz^2} = 4\pi G \rho_{\mathrm{tot}}(z),
    \label{eq:poisson}
\end{equation}
where $\rho_{\mathrm{tot}}(z)$ is the sum over mono-age isothermal subpopulations given by \equref{eq:rho_z}, representing stellar subpopulations together with the adopted gas, stellar-halo, and dark-matter density components. The resulting density profiles are therefore dynamically consistent with the derived potential, and through the AVR older disc subpopulations naturally acquire larger velocity dispersions, and hence larger scale heights, than younger ones.

The radial term in \equref{eq:poisson} is neglected as in a strongly flattened system such as the MW disc, radial gradients in density and gravitational potential are much smaller than vertical ones \citep{Just2010}. This assumption holds well for relatively low $z$ and intermediate $R$ considered in this work.

The model has been developed and calibrated incrementally. The original local version was calibrated using \textit{Hipparcos} data \citep{Perryman1997}, the Catalogue of Nearby Stars \citep{Jahreiss1997}, and kinematic constraints in the Solar cylinder \citep{Just2010}; SDSS star counts towards the North Galactic Pole were subsequently used to improve the SFH and the thick-disc parameters \citep{Just2011}, and a Bayesian calibration of the IMF based on \textit{Hipparcos} and CNS was introduced together with a more explicit treatment of stellar evolution in the population synthesis \citep{Rybizki2015}. Comparisons with Tycho--Gaia Astrometric Solution \citep[TGAS; ][]{Gaia2016} and Radial Velocity Experiment \citep[RAVE;][]{Steinmetz2006} then demonstrated that the model reproduces local star counts and kinematics at the several-per-cent level \citep{Sysoliatina2018}. Most recently, the radial extension of \citet{Sysoliatina2022} generalised the framework to $4 \lesssim R \lesssim 14$ kpc by treating the disc as a set of dynamically independent radial annuli.

Of particular relevance to the present work is the local calibration of \citetalias{Sysoliatina2021}, which optimised 22 model parameters against \textit{Gaia} DR2 and Apache Point Observatory Galactic Evolution Experiment \citep[APOGEE;][]{Bovy2014} red clump stars in a Bayesian framework and reproduced local star counts to about 5\%. That version introduced updated gas components, a time-extended thick-disc SFR, a four-slope broken-power-law IMF, and an AMR constrained by APOGEE red clump stars, and it found evidence for two recent episodes of enhanced star formation centred at approximately \mbox{0.5 Gyr} and \mbox{3 Gyr} ago. The properties of these recent star formation episodes are re-examined through the calibration presented later in this work.

We adopt and extend this framework because its vertical stellar densities are computed consistently with the total gravitational potential and because it includes dynamical heating through the AVR. This is particularly important for modelling white dwarfs, which span a wide range of ages and therefore probe populations with very different vertical scale heights. We retain the kinematic calibration established in \citetalias{Sysoliatina2021} and focus on photometric constraints to investigate the impact of the WD population on the inferred SFH. In \citetalias{Sysoliatina2021}, the model was calibrated using three complementary types of observational constraints: colour–magnitude distributions, vertical density profiles, and vertical kinematics of different stellar populations. Here, we keep the kinematic parameters fixed to their previously calibrated values, rather than simultaneously re-optimising them together with the SFH parameters. This allows us to extend the model to include WDs and explore a broader range of SFH prescriptions while keeping the number of free parameters manageable.

\subsection{Stellar and white dwarf evolution}
\label{sec:stellar_evolution}

Stellar populations before the WD phase are modelled using the PARSEC isochrone grid\footnote{We used our automated isochrone download tool \texttt{isotoolkit}, \url{https://github.com/askenja/isotoolkit}.} \citep{Bressan2012, Nguyen2022}, as in our previous works \citep{Sysoliatina2021, Sysoliatina2022}. These tracks are used for all non-degenerate stars and for estimating the WD progenitor lifetimes that enter the WD population synthesis. The WD phase is treated separately because standard isochrone grids do not extend continuously from the MS through the TP-AGB and post-AGB phases onto the WD cooling sequence, primarily owing to uncertainties in late-stage mass loss and remnant formation.

For the WD population we use and compare three independent sets of cooling sequences, the Montr\'{e}al models\footnote{\url{http://www.astro.umontreal.ca/~bergeron/CoolingModels}} \citep[also known as STELUM;][]{Bedard2020}; LPCODE\footnote{\url{https://evolgroup.fcaglp.unlp.edu.ar/modelos.html}} \citep{Althaus2010, Camisassa2019, Camisassa2021}; and BaSTI\footnote{\url{http://basti-iac.oa-abruzzo.inaf.it/wdmodels.html}} \citep{Salaris2010, Salaris2022}. Using multiple model families allows us to quantify systematic differences in WD evolution and to assess their impact on the predicted CMD morphology, luminosity function, and inferred SFH.

Modern WD cooling theory includes several processes that strongly affect the cooling rate. At high luminosities, energy losses are dominated by neutrino emission, whereas at intermediate stages, cooling proceeds gravothermally. At low luminosities, core crystallisation modifies the cooling rate through the release of latent heat and the gravitational energy associated with phase separation \citep{Althaus2010, Salaris2010, Garcia_Berro2010}. These processes introduce significant cooling delays, observed directly as features in the WD sequence \citep{Tremblay2019}. Further delays arise from the sedimentation of neutron-rich species such as $^{22}$Ne, particularly in metal-rich progenitors
\citep{Camisassa2021}. More recent studies have shown that $^{22}$Ne can also undergo distillation during crystallisation, whereby buoyant, $^{22}$Ne-depleted crystals redistribute neutron-rich material towards the core, releasing additional gravitational energy and potentially delaying cooling by several Gyr \citep{Blouin2021, Bauer2023, Bedard2024}. This process is not included in the three publicly available cooling grids adopted here. Its effects may be partially approximated by the phenomenological cooling-delay parameter introduced in \secref{Sec:ModelCalibration}, although this parameter cannot reproduce the expected mass- and luminosity-dependent behaviour of the additional delay.

The Montréal sequences provide extensive grids over a wide range of WD masses for both DA (hydrogen-atmosphere) and DB (helium-atmosphere) WDs. Their cooling sequences of \citet{Bedard2020} incorporate updated conductive opacities and include crystallisation, but neglect C/O phase separation and assume a homogeneous, equimass C/O core. The LPCODE sequences, by contrast, are based on continuous evolutionary calculations from the MS through the thermally pulsing AGB to the WD stage, yielding self-consistent core chemical profiles and envelope structures; they cover He-core, CO-core, and ONe-core WDs and include crystallisation, phase separation, and $^{22}$Ne sedimentation. The BaSTI models provide a third, independent set of CO-core cooling tracks based on alternative input physics and progenitor assumptions, again including crystallisation and phase separation.

Two properties of the BaSTI grid require comment. First, we adopt the BaSTI sequences with the electron conduction opacities of \citet{Blouin2020} and a combined metallicity grid, for DA white dwarfs the composition closest to solar that is available is $Z = 0.01$, which we also adopt for DB white dwarfs above \mbox{0.9 M$_\odot$}, whereas DB models in the range \mbox{0.54--0.9 M$_\odot$} are available only at $Z = 0$. Second, the high-mass limit of the BaSTI grid is \mbox{1.1 M$_\odot$}, lower than the \mbox{1.3 M$_\odot$} of the Montr\'{e}al and LPCODE grids, which introduces some incompleteness at the high-mass end.

The low-mass limit of all three grids is \mbox{0.54 M$_\odot$}, close to the usual lower limit of the IFMR applicability range. Lower-mass WDs are predominantly helium-core remnants formed through binary evolution \citep{Cummings2018}; since we do not model binary evolution channels, these are excluded.

Synthetic photometry is obtained by combining WD cooling sequences with model atmospheres and bolometric corrections. The Montréal sequences use the synthetic-colour tables of \citet{Holberg2006}, based on \citet{Bergeron1995}, with updated atmospheric calculations. BaSTI adopts bolometric corrections from the same atmospheric framework \citep{Salaris2010, Salaris2022}, whereas LPCODE uses the independent atmosphere models of \citet{Koester2010}, as described by \citet{Camisassa2021}. Thus, WD-BA and WD-MT primarily differ in their cooling prescriptions, while comparisons involving WD-LP also reflect differences in atmospheric modelling.

Although all three model families incorporate the same broad physical processes, they differ in their treatment of progenitor evolution, core composition stratification, envelope structure, and the implementation of crystallisation, phase separation, and sedimentation. These differences can produce measurable offsets in cooling times, particularly at faint magnitudes, and may propagate into systematic differences in the modelled WD number counts and the reconstructed SFH.

We interpolate all cooling sequences onto a uniform grid, constructing WD isochrones with a mass resolution of \mbox{0.002 M$_\odot$} and an age resolution of \mbox{50 Myr}. The total age
of a WD is approximated as
\begin{equation}
    \tau_\mathrm{WD} \approx t_\mathrm{MS} + t_\mathrm{cool},
    \label{eq:tau_wd}
\end{equation}
where $t_\mathrm{MS}$ is the progenitor's MS lifetime and $t_\mathrm{cool}$ the cooling time, neglecting the comparatively short post-MS phases that precede settling onto the cooling sequence.

\begin{figure}
\includegraphics[width=\columnwidth]{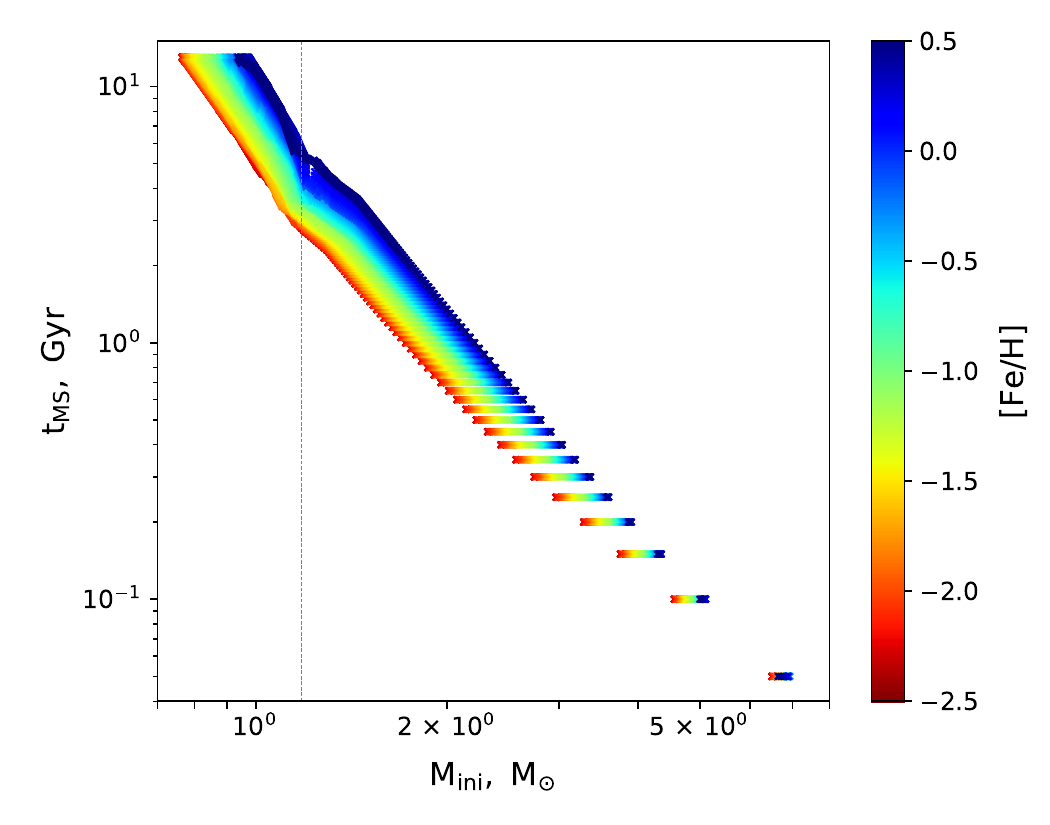}
\caption{MS lifetime, $t_\mathrm{MS}$, as a function of initial mass, $M_\mathrm{ini}$, from the PARSEC evolutionary tracks for the metallicity range considered in this work. The vertical grey dashed line marks the adopted break mass, $M_\mathrm{br}=1.18\,\mathrm{M}_\odot$, used in the piecewise linear approximation of the $\log t_\mathrm{MS}$--$\log M_\mathrm{ini}$ relation. The mean mass spacing of the PARSEC grid is $0.022\,\mathrm{M}_\odot$.}
\label{fig:t-ms}
\end{figure}

Main-sequence lifetimes are often approximated by analytic relations that neglect metallicity. Instead, we derive \mbox{$t_\mathrm{MS}(M_\mathrm{ini}, \mathrm{[Fe/H]})$} directly from the PARSEC isochrones. As shown in \figref{fig:t-ms}, metallicity affects stellar lifetimes significantly, with variations of up to a factor of two at fixed mass; neglecting this dependence can therefore bias the inferred total ages of WDs and distort the recovered SFH.

For each metallicity we approximate the relation in $\log t_\mathrm{MS}$--$\log M_\mathrm{ini}$ space using a two-slope piecewise linear model with a fixed break mass $M_\mathrm{br} = 1.18\,\mathrm{M}_\odot$. For $\mathrm{[Fe/H]} \lesssim 0$ and $M_\mathrm{ini} < M_\mathrm{br}$ the slope is close to the classical value of $-2.5$, while higher-mass and more metal-rich stars show a steeper dependence. We construct a bilinear interpolator, $\mathcal{T}$, on this grid,
\begin{equation}
\label{eq:t_ms}
    \log t_{\mathrm{MS}} = \mathcal{T}\left(\mathrm{[Fe/H]},\,
    \log M_\mathrm{ini}\right),
\end{equation}
where the progenitor mass follows from the IFMR\footnote{We parameterise the forward relation and numerically invert it to obtain the progenitor mass.},
\begin{equation}
    M_\mathrm{ini} = \mathrm{IFMR}(M_\mathrm{f}).
\end{equation}

The resulting age--metallicity WD isochrones are appended to the corresponding PARSEC isochrones. Metallicity therefore enters the WD population only through progenitor lifetimes and initial masses, since the cooling sequences adopted here are computed at near solar composition.

\subsection{Stellar and white dwarf population synthesis}
\label{sec:stellar_and_wd_popsynth}

To convert the modelled density profiles (\secref{sec:model_basic}) into mock stellar populations on the CMD, we use the joint PARSEC$+$WD isochrones described in \secref{sec:stellar_evolution}, which map age--metallicity pairs to observable properties. We note that the choice of stellar library can significantly affect the predicted CMD morphology introducing systematic uncertainties.

We introduce a Gaussian metallicity spread of \mbox{0.07 dex} \citepalias[][]{Sysoliatina2021} around each age--metallicity pair of the AMR. This accounts for the intrinsic scatter of the AMR and to some extent approximates the effects of radial migration. Each mono-age subpopulation is therefore split into seven mono-metallicity subpopulations sampling the mean AMR value and $\pm1\sigma$, $\pm2\sigma$, and $\pm3\sigma$ from it.

For each mono-(age--metallicity) subpopulation, the corresponding isochrone defines a set of stars of different initial masses. Following our previous works, we refer to these mono-(age--metallicity--mass) groups as stellar assemblies (SAs). The number of SAs per subpopulation is set by the mass grid of the PARSEC library and by the adopted mass resolution of the interpolated WD isochrones. The present-day SA surface number densities, $N_\mathrm{j}^\Sigma$, follow from the IMF, the SFH, and the integrated stellar mass loss, which together determine the number of stars formed at a given epoch that survive to
the present day.

These surface densities are converted into vertical number density distributions assuming dynamical equilibrium in the Galactic potential, using the AVR and the total vertical potential provided by the model. The total number of stars per stellar assembly in the modelled volume is
\begin{equation}
N_\mathrm{j}^V = \frac{N_\mathrm{j}^\Sigma}{2\,h(\tau_\mathrm{j})} 
\sum_k \exp\left[-\frac{\Phi(|z_\mathrm{k}|)}{\sigma_W^2(\tau_\mathrm{j})}\right] \cdot V(z_\mathrm{k}, M_\mathrm{G,\,j}),
\label{eq:Nzv}
\end{equation}
where $h(\tau_\mathrm{j})$ is the effective scale height of the subpopulation, which normalises the vertical density profile given by \equref{eq:rho_z}. The effective volume $V(|z_\mathrm{k}|, M_\mathrm{G,j})$ depends on both height above the plane and the SA absolute magnitude, since the observational selection imposes magnitude-dependent distance limits (see \secref{sec:completeness} and \figref{fig:cmd_completeness}). This ensures that the model reproduces the magnitude-limited nature of the observed sample.

The same framework can be extended beyond the stellar regime by coupling the Galactic population synthesis to substellar evolutionary models. We demonstrate this capability in \secref{sec:results_bd} using the models from \citet{Phillips2020_ATMO}.

\section{Model calibration}
\label{Sec:ModelCalibration}

In this section, we describe the free parameters explored with the Markov chain Monte Carlo (MCMC) sampling, the computationally efficient scheme used to generate CMD predictions from each model realisation, and the likelihood through which the model is compared to the data.

\subsection{Free parameters in the MCMC}

We sample 18 free parameters, collected in the vector $\theta$. Below we describe each group and the corresponding prior; all priors are listed in \tabref{tab:best_param}. Throughout, MCMC exploration is restricted to within $3\sigma$ of each prior mean.

\textit{IFMR parameters}: We model the IFMR as a three-slope piecewise-linear function and vary its shape through an offset $b_0$ of the lowest-mass segment, three slopes $\alpha^\mathrm{IFMR}_0$, $\alpha^\mathrm{IFMR}_1$, $\alpha^\mathrm{IFMR}_2$, and two break masses $m^\mathrm{IFMR}_0$, $m^\mathrm{IFMR}_1$. Prior means are taken from the PARSEC-calibrated solution of \citet{Cummings2018} \citepalias[hereafter][]{Cummings2018}, with slope dispersions equal to the \citetalias{Cummings2018} values, rounded. We deliberately adopt wide dispersions for the offset, $\Delta b_0 = 0.1$, and for the break masses, $\Delta m^\mathrm{IFMR}_0 = \Delta m^\mathrm{IFMR}_1 = 0.14$ M$_\odot$, the latter being the largest value consistent with $m^\mathrm{IFMR}_1 > m^\mathrm{IFMR}_0$.

\textit{WD DA/DB fraction}: Although the observed DA/DB ratio, $f_\mathrm{DA}$, varies with effective temperature as a result of spectral evolution \citep{Bedard2020, LopezSanjuan2022}, we neglect this dependence and instead assume a single, temperature-independent DA fraction throughout the WD population \citep[e.g.,][]{Tremblay2008}. This constant fraction is treated as a free parameter in the MCMC calibration, allowing the data to determine the effective population-averaged DA fraction. While this approximation is appropriate for the present population-synthesis model, future extensions could incorporate either an empirical temperature-dependent DA fraction or a self-consistent treatment of spectral evolution.
    
\textit{WD cooling delay}: To absorb residual uncertainty in the cooling sequences, we introduce a parameter $\alpha_\mathrm{cool}$ that applies a fractional slowdown to the cooling time (\equref{eq:d_cool}), acting on both DA and DB isochrones. The prior is centred on no delay, $\langle \alpha_\mathrm{cool}\rangle = 0$, with $\Delta \alpha_\mathrm{cool} = 0.1$ that corresponds to a \mbox{10\%} cooling slowdown. This parameter accounts for systematic offsets between the adopted cooling models and the observed WD population, including both uncertainties in the cooling physics and the approximation of neglecting the short post-MS evolutionary phases when computing WD ages.

\textit{Thin-disc SFH parameters}: The overall shape of the thin-disc SFH is controlled by $\eta$ and $\zeta$, which govern how sharply peaked or shallow the continuum is. Their prior means are taken from \citetalias{Sysoliatina2021}, and the dispersions $\Delta\eta$, $\Delta\zeta$ are chosen wide enough that the SFR can range from strongly old-age-peaked and rapidly declining, through constant (the case $\zeta = \eta/2$, see equations~10--12 of \citetalias{Sysoliatina2021}), to rising with cosmic time. This deliberately broad prior lets us test SFH shapes very different from our default declining form, including those reported in WD-based studies. We additionally vary the SFH normalisation $\Sigma_\mathrm{d}$ and two parameters of a secondary star-formation episode, its amplitude $\Sigma_\mathrm{p1}$ and mean age $\tau_\mathrm{p1}$, with priors consistent with \citetalias{Sysoliatina2021}; the dispersions allow this episode to be removed entirely or shifted from $\langle\tau_\mathrm{p1}\rangle = 3.5$ Gyr to the present. In addition, only the amplitude of the most recent star formation episode is treated as a free parameter.
    
\textit{IMF parameters}:  The IMF is explored through three slopes $\alpha^\mathrm{IMF}_0$, $\alpha^\mathrm{IMF}_1$, $\alpha^\mathrm{IMF}_2$ and two break masses $m^\mathrm{IMF}_0$, $m^\mathrm{IMF}_1$, with priors from \citetalias{Sysoliatina2021}. The fourth (high-mass) slope above \mbox{6 M$_\odot$} remains fixed, as the local sample does not constrain such rare, short-lived stars.

\subsection{Generation of the predicted CMD}

The SA tables are computed once for a reference vector $\theta_0$. When the parameters change, the Poisson equation (\equref{eq:poisson}) is re-solved for the new potential--density pair, and the SA surface densities are updated by re-weighting their reference values. Each weight corrects the reference density for the effect of a particular parameter change; we give the individual terms below. We mark quantities computed from the updated parameters with a prime and leave reference quantities unmarked, and we drop the SA index $j$ for brevity.

\subsubsection{Main-sequence and giant populations}

For the MS and giant stars, the only varied inputs are the SFR and IMF, while the total age and initial mass of each SA are unchanged. Two corrections are therefore required, accounting for the change in the SFR and IMF probabilities in the relevant age and mass bins:
\begin{subequations}
\begin{align}
\omega_\mathrm{SFR}(\tau) &=
\mathrm{SFR}'(\tau)/\mathrm{SFR}(\tau),
\\
\omega_\mathrm{IMF}(M_\mathrm{ini}) &=
\mathrm{IMF}'(M_\mathrm{ini})/\mathrm{IMF}(M_\mathrm{ini}).
\end{align}
\end{subequations}
The combined weight for non-degenerate populations is
\begin{equation}
    \omega_{\mathrm{non\text{-}WD}} = \omega_{\mathrm{SFR}}(\tau) \cdot
    \omega_{\mathrm{IMF}}(M_\mathrm{ini}).
    \label{eq:w_noWD}
\end{equation}

\subsubsection{White dwarf populations}

For WDs, several updates precede the analogue of  \equref{eq:w_noWD}, because a change in the IFMR or cooling delay alters both the progenitor mass and the total age of a given SA. First, the cooling time is rescaled by the cooling delay parameter,
\begin{equation}
    t'_{\mathrm{cool}} = (1 + \alpha'_{\mathrm{cool}})\, t_{\mathrm{cool}}.
    \label{eq:d_cool}
\end{equation}
Second, the progenitor mass is updated through the new IFMR, $M'_\mathrm{ini} = \mathrm{IFMR}'(M_\mathrm{f})$, the MS lifetime is recomputed with the interpolator of \equref{eq:t_ms} at the new mass and unchanged metallicity, and the updated total age $\tau'$ then follows from \equref{eq:tau_wd}.

Because the progenitor mass and total age have both shifted, a given SA is now mapped to different mass and age bins of the reference IMF and SFR. We account for this relocation with a shift weight,
\begin{equation}
    \omega_\mathrm{shift}(M'_\mathrm{ini}, \tau') =
    \frac{\mathrm{IMF}(M'_\mathrm{ini})}{\mathrm{IMF}(M_\mathrm{ini})}
    \cdot
    \frac{\mathrm{SFR}(\tau')}{\mathrm{SFR}(\tau)},
    \label{eq:w_shift}
\end{equation}
in which the reference (unprimed) IMF and SFR are evaluated at the new and old arguments, the first factor relocates the SA in initial mass and the second in age, both under the reference model. The updated parameters then enter separately, through the reweighting factors $\omega_\mathrm{SFR}(\tau')$ and $\omega_\mathrm{IMF}(M'_\mathrm{ini})$ defined as in \equref{eq:w_noWD} but evaluated at the shifted age and mass.

The change in atmospheric fraction is applied through
\begin{equation}
\omega_\mathrm{DX} =
\begin{cases}
f'_{\mathrm{DA}}/f_{\mathrm{DA}}, & \text{DA white dwarfs}, \\[2pt]
(1 - f'_{\mathrm{DA}})/(1 - f_{\mathrm{DA}}), & \text{DB white dwarfs}.
\end{cases}
\end{equation}
The total WD weight is the product of the relocation, atmospheric, and reweighting factors,
\begin{equation}
    \omega_{\mathrm{WD}} = \omega_\mathrm{shift}(M'_\mathrm{ini}, \tau')
    \cdot \omega_\mathrm{DX}
    \cdot \omega_{\mathrm{SFR}}(\tau')
    \cdot \omega_{\mathrm{IMF}}(M'_\mathrm{ini}),
    \label{eq:w_WD}
\end{equation}
and the updated SA densities are
\begin{equation}
N' =
\begin{cases}
 N \cdot \omega_\mathrm{WD}, & \text{white dwarfs}, \\[2pt]
 N \cdot \omega_{\mathrm{non\text{-}WD}}, & \text{other populations}.
\end{cases}
\end{equation}
At each iteration these densities are binned into the Hess diagram, and projected onto the colour and magnitude distributions of the CMD regions (\equref{eq:CMD_cuts}) that are compared to the data as described below.

\subsection{Bayesian inference}
\label{sec:Inference}
We infer the model parameters within a Bayesian framework, expressing the posterior $\mathbb{P}$ through the prior $P$ and the likelihood $\mathcal{L}$ in logarithmic form,
\begin{equation}
    \log \mathbb{P} = \log \mathcal{L} + \log P,
\end{equation}
and adopting an independent Gaussian prior on each parameter,
\begin{equation}
    \log P = -0.5 \sum_\mathrm{s}
    \frac{(\theta_\mathrm{s} - \theta_{0,\mathrm{s}})^2}{\sigma_\mathrm{s}^2},
    \label{eq:prior}
\end{equation}
where $\theta_{0,\mathrm{s}}$ and $\sigma_\mathrm{s}$ are the prior mean and dispersion of the $s$-th parameter (\tabref{tab:best_param}).

The parameter constraints come from the likelihood, which measures the consistency between the modelled and observed Hess diagrams. Fitting a Hess diagram is not straightforward; it is a two-dimensional distribution with strong density gradients and many empty or noise-dominated bins, and the stellar evolution models reproduce several CMD features, i.e., the giant branch, the blue edge of the UMS, the lower-MS slope, and the overall MS shape, only with systematic offsets and distortions relative to the data. A naive bin-by-bin $\chi^2$ over the full diagram therefore has several drawbacks. The few high-density regions would dominate the likelihood, leaving sparse but physically important loci such as the giant branch and the WD sequence effectively unconstrained. The systematic offsets between the modelled and observed high-density regions mean that a bin-by-bin match does not necessarily identify the best global parameter combination. And suppressing the many nearly empty bins would require either additional hyperparameters (a count offset or a minimum occupancy threshold) or the inclusion of nuisance parameters in the sampling, the latter being more rigorous but costly in convergence and run time.

To avoid these pitfalls, we divide the Hess diagram into the five CMD regions defined in \equref{eq:CMD_cuts} and shown in \figref{fig:data}, and require each region to contribute with equal weight to the likelihood. In this way the fit responds simultaneously to the dense MS and to the sparser WD and giant populations. Within each region we further compare one-dimensional projections of the Hess diagram rather than the full two-dimensional distribution, which stabilises the MCMC convergence and greatly reduces the influence of empty or noise-dominated bins. 

The likelihood is constructed as a sum over the five regions $l \in \{\mathrm{G, UMS, MS, LMS, WD}\}$ of three terms,
\begin{equation}
\log \mathcal{L}
= \sum_{l}
\left(
\log \mathcal{L}^{\mathrm{CD}}_{\mathrm{shape},l}
+ \log \mathcal{L}^{\mathrm{MD}}_{\mathrm{shape},l}
+ \log \mathcal{L}_{\mathrm{count},l}
\right),
\label{eq:loglike}
\end{equation}
two of which constrain the shapes of the region's area-normalised colour (CD) and absolute magnitude (MD) distributions, and one of which constrains its total star count. We note that these three terms are derived from the same stars and are therefore not statistically independent; \equref{eq:loglike} is thus best understood as an effective goodness-of-fit metric, designed to weight the five CMD regions comparably, rather than as a formal joint likelihood. As we show below, the systematic floors that enter each term dominate its variance, so that the location of the posterior maximum is set by the relative residuals across regions and is insensitive to the neglected correlations. The marginal parameter uncertainties in \tabref{tab:best_param} should correspondingly be read as internal precisions of the fit rather than as fully calibrated statistical errors.

We define the two shape terms as binned $\chi^2$ statistics,
\begin{equation}
\log \mathcal{L}^{X}_{\mathrm{shape},l}
= -0.5 \sum_\mathrm{m}
\frac{(D_\mathrm{m} - M_\mathrm{m})^2}{\sigma_\mathrm{m}^2}\, \Delta_X,
\qquad X \in \{\mathrm{CD, MD}\},
\label{eq:like_shape}
\end{equation}
where $m$ indexes the bins of the one-dimensional distribution, $D_\mathrm{m}$ and $M_\mathrm{m}$ are the normalised data and model values, $\sigma_\mathrm{m}$ is the per-bin uncertainty, and $\Delta_X$ is the bin width. Because the colour and magnitude distributions are binned at different resolutions, weighting each term by $\Delta_X$ keeps the total shape contribution insensitive to the choice of binning and places the two projections on a comparable footing, so that neither dominates the shape likelihood by virtue of being more finely sampled.

The per-bin variance combines a Poisson term with a constant systematic floor,
\begin{equation}
\sigma_\mathrm{m}^2 = \frac{D_\mathrm{m}}{N_\mathrm{D,l}}
+ \sigma^2_{\mathrm{CMD}},
\label{eq:sigma_shape}
\end{equation}
where $N_\mathrm{D,l}$ is the total number of stars in region $l$. The floor $\sigma^2_{\mathrm{CMD}}$ accounts for intrinsic model imperfections and prevents the fit from over-reacting to small, systematically driven discrepancies. Given the sample size, $N_\mathrm{D,l} \sim 10^{5}$--$10^{6}$ per region, the Poisson term is negligible relative to this floor, so the shape likelihood is governed almost entirely by the systematic component.

For these shape terms we use the cleaned subsample $S_1$ (\secref{Sec:Data}), since unresolved binaries are not included when converting the stellar assemblies into Hess-diagram counts, and we therefore remove them from the data as well so as to compare only the single-star CMD footprints. Removing binaries measurably affects only the faint LMS and WD distributions and has a negligible effect on the G, UMS, and MS shapes. Because only the normalised shapes are compared, the absolute completeness of $S_1$ does not enter these terms.

\setlength{\extrarowheight}{0.5em}
\begin{table*}
\centering
\caption{Posterior estimates of the free parameters obtained from the MCMC optimisation for the three adopted white dwarf cooling models described in \secref{sec:stellar_evolution}, LPCODE (WD-LP), BaSTI (WD-BA), and Montréal (WD-MT). Quoted uncertainties correspond to the 16\textsuperscript{th} and 84\textsuperscript{th} percentiles of the marginal posterior distributions for all parameters, except $\alpha_\mathrm{cool}$ that has a truncated prior distribution. For $\alpha_\mathrm{cool}$, we report the value of the posterior distribution's mode with the uncertainties calculated using a bootstrapping method. The final column lists a relative shrinkage of the width of each parameter probability distribution with respect to its prior width, $S_{\rm rel}$.}
\begin{tabular}{cc|l|ccccc} 
\hhline{========}
Parameter & Units & Description & Prior
& {\begin{minipage}[t][0.4cm][t]{1.5cm} \center{WD-LP} \end{minipage}}
& {\begin{minipage}[t][0.4cm][t]{1.5cm} \center{WD-BA} \end{minipage}} 
& {\begin{minipage}[t][0.4cm][t]{1.5cm} \center{WD-MT} \end{minipage}} 
& $\mathrm{S_{rel}}$
\\ 
\hhline{--------}
$b_{0}$ & M$_\odot$ & IFMR offset for M$_\mathrm{ini} \in (0.87,m^\mathrm{{IFMR}}_0)$ M$_\odot$ & $0.476\pm0.1^{\,(1)}$ & 
$0.508^{+0.019}_{-0.019}$ & 
$0.504^{+0.018}_{-0.018}$ & 
$0.51^{+0.018}_{-0.019}$ &
0.62 \\
$\alpha^\mathrm{{IFMR}}_{0}$ & & IFMR slope for M$_\mathrm{ini} \in (0.87,m^\mathrm{{IFMR}}_0)$ M$_\odot$ & $0.0873\pm0.02^{\,(1)}$ & 
$0.098^{+0.019}_{-0.018}$ & 
$0.096^{+0.019}_{-0.018}$ & 
$0.098^{+0.019}_{-0.018}$ &
0.08 \\
$\alpha^\mathrm{{IFMR}}_{1}$ & & IFMR slope for M$_\mathrm{ini} \in (m^\mathrm{{IFMR}}_0,m^\mathrm{{IFMR}}_1)$ & $0.181\pm0.04^{\,(1)}$ & 
$0.181^{+0.04}_{-0.04}$ & 
$0.181^{+0.04}_{-0.039}$ & 
$0.181^{+0.039}_{-0.039}$ &
0 \\
$\alpha^\mathrm{{IFMR}}_{2}$ & & IFMR slope for M$_\mathrm{ini} \in (m^\mathrm{{IFMR}}_1,8.2)$ M$_\odot$ & $0.0835\pm0.015^{\,(1)}$ & 
$0.0843^{+0.014}_{-0.015}$ & 
$0.0838^{+0.014}_{-0.015}$ & 
$0.0842^{+0.0137}_{-0.0145}$ &
0.03 \\
$m^\mathrm{{IFMR}}_{0}$ & M$_\odot$ & 1$^\text{st}$ IFMR break point & $2.8\pm0.14^{\,(1)}$  & 
$2.8^{+0.14}_{-0.14}$ & 
$2.81^{+0.14}_{-0.14}$ & 
$2.8^{+0.14}_{-0.14}$ &
0 \\
$m^\mathrm{{IFMR}}_{1}$ & M$_\odot$ & 2$^\text{nd}$ IFMR break point & $3.65\pm0.14$  & $3.65^{+0.14}_{-0.14}$ & 
$3.65^{+0.14}_{-0.14}$ & 
$3.65^{+0.14}_{-0.14}$ &
0 \\
$\alpha_\mathrm{cool}$ & & WD cooling delay & $|0\pm0.1|$ & 
$0.017^{+0.001}_{-0.001}$ & 
$0.011^{+0.001}_{-0.001}$ & 
$0.001^{+0.0003}_{-0.0003}$ &
-0.05 \\
$f_\mathrm{DA}$ & & WD DA fraction & $0.8\pm0.065$ & 
$0.77^{+0.06}_{-0.06}$ & 
$0.77^{+0.06}_{-0.06}$ & 
$0.78^{+0.06}_{-0.07}$ &
0.08 \\[5pt]\cdashline{1-8}
$\zeta$ & & $\mathrm{SFR_d}$ shape parameter & 
$0.83\pm0.27^{\,(2)}$ & 
$0.58^{+0.2}_{-0.2}$ & 
$0.56^{+0.2}_{-0.21}$ & 
$0.56^{+0.21}_{-0.21}$ &
0.26 \\ 
$\eta$ & & $\mathrm{SFR_d}$ shape parameter & 
$5.6\pm1.5^{\,(2)}$ & 
$5.02^{+0.99}_{-0.98}$ & 
$4.97^{+1.01}_{-0.99}$ & 
$4.89^{+1.04}_{-1.02}$ &
0.34 \\
$\Sigma_\mathrm{d}$ & M$_\odot$ pc$^{-2}$ & Thin-disc surface density & $29.4\pm3^{\,(2)}$ & 
$31.2^{+1.8}_{-1.9}$ & 
$31.3^{+1.8}_{-2}$ & 
$31.6^{+1.8}_{-2}$ & 
0.38 \\
$\Sigma_\mathrm{p1}$ & M$_\odot$ pc$^{-2}$ & Relative amplitude of the 1$^\text{st}$ $\mathrm{SFR_d}$ burst & $3.5\pm 1.15^{\,(2)}$ & 
$3.44^{+1.16}_{-1.15}$ & 
$3.43^{+1.13}_{-1.14}$ & 
$3.51^{+1.14}_{-1.13}$ & 
0 \\
$\tau_\mathrm{p1}$ & Gyr & Mean age of the 1$^\text{st}$ $\mathrm{SFR_d}$ burst & 
$3.0\pm1.0^{\,(2)}$ & 
$2.42^{+1.04}_{-0.78}$ & 
$2.35^{+1.04}_{-0.78}$ & 
$2.45^{+1.04}_{-0.79}$ &
0.08 \\[5pt]\cdashline{1-8}
$\alpha^\mathrm{{IMF}}_{0}$ & & IMF slope for $M_\mathrm{ini} \in (0.08,m^\mathrm{{IMF}}_0)$ M$_\odot$ & 
$1.31\pm0.25^{\,(2)}$ & 
$1.04^{+0.18}_{-0.13}$ & 
$1.04^{+0.17}_{-0.13}$ & 
$1.05^{+0.18}_{-0.13}$ & 
0.38 \\
$\alpha^\mathrm{{IMF}}_{1}$ & & IMF slope for $M_\mathrm{ini} \in (m^\mathrm{{IMF}}_0,m^\mathrm{{IMF}}_1)$ & 
$1.43\pm0.25^{\,(2)}$ & 
$1.61^{+0.08}_{-0.07}$ & 
$1.6^{+0.08}_{-0.08}$ & 
$1.62^{+0.08}_{-0.08}$ & 
0.7 \\
$\alpha^\mathrm{{IMF}}_{2}$ & & IMF slope for $M_\mathrm{ini} \in (m^\mathrm{{IMF}}_1,6)$ M$_\odot$ & 
$2.88\pm0.25^{\,(2)}$ & 
$2.94^{+0.19}_{-0.2}$ & 
$2.94^{+0.19}_{-0.2}$ & 
$2.9^{+0.19}_{-0.2}$ & 
0.22 \\
$m^\mathrm{{IMF}}_{0}$ & M$_\odot$ & 1$^\text{st}$ IMF break point & $0.49\pm0.15^{\,(2)}$  & 
$0.5^{+0.07}_{-0.09}$ & 
$0.5^{+0.07}_{-0.09}$ & 
$0.5^{+0.08}_{-0.1}$ & 
0.47 \\
$m^\mathrm{{IMF}}_{1}$ & M$_\odot$ & 2$^\text{nd}$ IMF break point & $1.43\pm0.15^{\,(2)}$  & 
$1.5^{+0.11}_{-0.12}$ & 
$1.5^{+0.11}_{-0.12}$ & 
$1.5^{+0.12}_{-0.12}$ & 
0.23 \\[5pt]
\hhline{========}
\end{tabular} 
\begin{tablenotes}
 \item References. (1) \citetalias{Cummings2018}; 
 (2) \citetalias{Sysoliatina2021}. 
\end{tablenotes}
\label{tab:best_param}
\end{table*}

The third term compares the total star counts,
\begin{equation}
\log \mathcal{L}_{\mathrm{count},l}
= -0.5\,
\frac{(N_\mathrm{D,l} - N_\mathrm{M,l})^2}{\sigma^2_\mathrm{N,l}},
\qquad
\sigma^2_\mathrm{N,l} = N_\mathrm{M,l} + (\epsilon\, N_\mathrm{M,l})^2,
\label{eq:like_count}
\end{equation}
where $N_\mathrm{D,l}$ and $N_\mathrm{M,l}$ are the observed and modelled star counts in region $l$. As in \equref{eq:sigma_shape}, the variance combines a Poisson term with a systematic term, here parametrised by a single hyperparameter $\epsilon$ that represents the expected fractional model-to-data count mismatch. We adopt $\epsilon = 0.02$ systematic uncertainty for all regions. This term is evaluated on the full sample $S_0$, so that it constrains the true total number of stars in the studied volume. As with the shape floor, the systematic term dominates the Poisson term for all regions, so $\epsilon$ effectively sets the weight of the count constraint.

We performed several validation runs with test parameter sets to ensure that this likelihood behaves as intended, the shape and count contributions scale with the observed discrepancies, empty and noise-dominated bins are suppressed, and the five CMD regions contribute comparably to the total. The typical likelihood and prior values are of the same order, with the prior beginning to dominate only when several parameters deviate from their means by more than $2\sigma$. Sampling the posterior with MCMC and locating its maximum therefore yields a well-behaved, reproducible parameter estimate.

\subsection{MCMC setup and simulations}

We sample the posterior with the affine-invariant ensemble sampler implemented in \texttt{emcee} \citep{foreman13}, whose parallel structure distributes the computation across multiple cores; the present analysis used 42 CPU cores. The 18-dimensional parameter space is explored with 72 walkers, four per parameter. 

The walkers are initialised by sampling from a multivariate normal distribution centred on $\theta_\mathrm{ini}$, with independent Gaussian perturbations of standard deviation $10^{-4}\,\Delta\theta$ applied to each parameter, where $\Delta\theta$ is the vector of prior dispersions. Because our default model, which adopts the \citetalias{Sysoliatina2021} parameters, substantially mismatches the observed WD counts (\secref{sec:results_sj21}), $\theta_\mathrm{ini}$ is not identical to the prior means $\theta_0$; we instead obtained it from a set of short exploratory runs and adopted their typical values. This produces a compact ensemble near the expected solution while retaining stochastic variation between walkers.

We ran three otherwise identical calibration simulations, differing only in the WD cooling grid used for the WD population synthesis, and refer to them as WD-LP (LPCODE), WD-BA (BaSTI), and WD-MT (Montr\'{e}al).

Each run comprised $2.1 \times 10^4$ iterations, chosen to satisfy $n_\mathrm{iter} \geq 50\,\hat\tau$, where $\hat\tau$ is the integrated autocorrelation time; this ensures that the chains extend well beyond the autocorrelation scale. Convergence was further assessed by visual inspection of the walker trajectories, confirming good mixing and the absence of trapped chains, and by monitoring the mean acceptance fraction, which reached $a_\mathrm{f} \approx 0.25$, consistent with efficient sampling.

For the posterior analysis we discarded a burn-in of $2\hat\tau$ iterations, using the mean autocorrelation time over the full chain, and thinned the remainder by selecting every $(\hat\tau/2)$-th sample. This yielded $\sim 7 \times 10^3$ approximately independent samples, from which we derived the marginal posterior distributions and the optimal parameters listed in \tabref{tab:best_param}. Quoted uncertainties correspond to the 16th and 84th percentiles of the marginal posteriors, and the relative shrinkage $S_\mathrm{rel} = 1 - \sigma_\mathrm{post}/\sigma_\mathrm{prior}$ in the final column measures how much each parameter is constrained relative to its prior.

\section{Results}
\label{Sec:Results}

\subsection{Predictions of the legacy SJ21 model}
\label{sec:results_sj21}

As a first step, we assess how well the original \citetalias{Sysoliatina2021} calibration reproduces the selected \textit{Gaia} sample, focusing on the two CMD regions newly added in this work, that is, the WD sequence and the LMS. This legacy model serves as a reference benchmark against which the recalibration is measured.

For this test, all free parameters are fixed to their prior values (\tabref{tab:best_param}), which either follow directly from \citetalias{Sysoliatina2021} or are set by the adopted WD physics; the \citetalias{Cummings2018} IFMR combined with the metallicity-dependent MS lifetimes from PARSEC (\figref{fig:t-ms}), a DA fraction of 0.8, no cooling delay, and the LPCODE cooling sequences. We refer to this realisation as SJ21-LP.

As expected, SJ21-LP reproduces the CMD regions on which it was originally calibrated reasonably well, the total star count is matched to $-1.7$\%, and the G, UMS, MS, and LMS counts to within about 10\% (\tabref{tab:counts-4models}, \figref{fig:hess-4models}). Its extension to the WD population, however, fails badly. The model overpredicts the total number of WDs by 81\%, with the largest excess among the faint, old WDs and a further systematic excess along the bright edge of the cooling sequence. The LMS total agrees to $-0.7$\%, but the Hess-diagram residuals reveal systematic discrepancies across most faint, low-mass bins.

This 81\% WD excess is the central motivation for the present work; a model calibrated on the main-sequence population alone does not provide an adequate description of the white dwarfs it implies. We therefore recalibrate the model using the WD population as an additional constraint used simultaneously with other population types.

\setlength{\extrarowheight}{0.25em}
\begin{table}
    \centering
    \small
    \setlength{\tabcolsep}{4pt}
    \begin{tabular}{llllllll}
    \toprule
     & \multicolumn{1}{c}{G} & \multicolumn{1}{c}{UMS} & \multicolumn{1}{c}{MS} 
     & \multicolumn{1}{c}{LMS} & \multicolumn{1}{c}{WD} & \multicolumn{1}{c}{Total} & \multicolumn{1}{c}{$\log \mathcal{L}$} \\
    \midrule
    SJ21-LP & $-5.9$ & $+7.8$ & $+10$ & $-0.7$ & $+81.4$ & $-1.7$ & $-312.4$ \\
    WD-LP & $-1.4$ & $+2.6$ & $+0.1$ & $+0.6$ & $+0.5$ & $+0.5$ & $-34.4$ \\
    WD-BA & $-1.3$ & $+1.8$ & $-0.2$ & $+0.04$ & $+0.8$ & $+0.1$ & $-33.6$ \\
    WD-MT & $-1.5$ & $+2.9$ & $+0.3$ & $+1.6$ & $-0.3$ & $+1.2$ & $-36.3$\\
    \bottomrule
    \end{tabular}
    \normalsize
    \caption{Fractional difference $N_\mathrm{m}/N_\mathrm{d} - 1$ given in percent for the five defined CMD regions (see  Table~\ref{tab:StarCountsS0S1}). Statistics were calculated for the legacy \citetalias{Sysoliatina2021} model with the LPCODE-based WD isochrones, SJ21-LP, and for the three realisations with the different WD cooling models and the corresponding optimised parameter sets from Table~\ref{tab:best_param}. The last column lists the corresponding logarithmic likelihood values.} 
    \label{tab:counts-4models}
\end{table}

\subsection{Model after calibration}
\label{sec:calibrated_model}

\subsubsection{The new parameters}
\label{sec:new_params}

\begin{figure*}
\includegraphics[width=\textwidth]{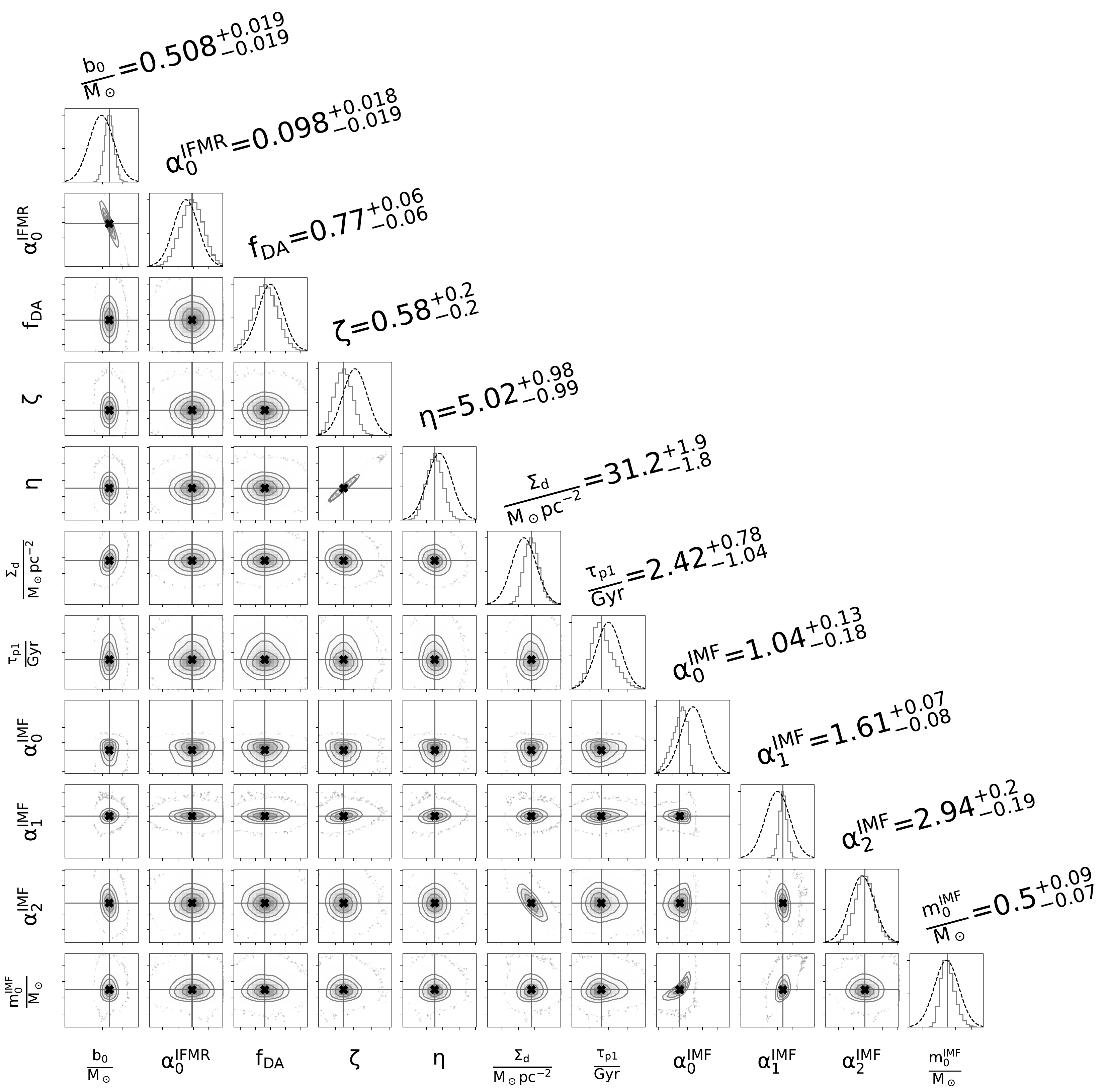}
\caption{Reduced corner plot for the WD-LP calibration. The diagonal panels show the one-dimensional posterior distributions of the main fitted parameters, while the off-diagonal panels show their joint posterior distributions. Dashed lines indicate the median and \mbox{16\textsuperscript{th}--84\textsuperscript{th}} percentile intervals.}
\label{fig:reduced_corner}
\end{figure*}

The optimised parameters from the WD-LP, WD-BA, and WD-MT runs are listed in \tabref{tab:best_param}. The inferred parameters agree across the three cooling prescriptions to well within their $1\sigma$ uncertainties, indicating that the recovered Galactic properties are largely insensitive to the adopted WD cooling model. We adopt WD-LP as the fiducial calibration as it yields marginally better agreement with the observational data. All predictions presented below in \secref{sec:PredictionOfModel} therefore use the WD-LP calibration. 

The posterior distributions of a representative subset of the WD-LP parameters are shown in \figref{fig:reduced_corner}, while the full posterior distribution including all fitted parameters is presented in \mbox{Appendix~\figref{fig:FullCornerPlot}}. 

Not all of the chosen free parameters can be robustly constrained by our modelling. In order to distinguish between the prior-dominated parameters and those for which the data provided some additional information, we defined a simple metrics $S_\mathrm{rel}$ (the last column of \tabref{tab:best_param}). It is a measure of the relative change in the parameter posterior width relative to the width of its prior. For the $s$-parameter, this quantity is defined as $S_\mathrm{rel,\,s} = 1 - \sigma_\mathrm{\mathbb{P},\,s}/\sigma_\mathrm{s}$. Here $\sigma_\mathrm{\mathbb{P},\, s}$ is the width of the MCMC-sampled posterior (diagonal in \figref{fig:reduced_corner} and \figref{fig:FullCornerPlot}). In general case, it is not a Gaussian distribution, so we calculate its width as $\sigma_\mathrm{\mathbb{P},\,s} = (\mathbb{P}_\mathrm{s}(<84\%) - \mathbb{P}_\mathrm{s}(<16\%))/2$. 

In the case of IFMR, only two parameters were strongly and reliably constrained: the offset $b_0$ and the first, lowest-mass, slope $\alpha^\mathrm{IFMR}_0$. Other IFMR slopes and the two mass break points were fully prior-dominated. The introduced WD cooling delay parameter, $\alpha_\mathrm{cool}$, has even somewhat wider posterior distribution than the input prior (negative $S_\mathrm{rel}$). Taking into account extremely small values obtained for $\alpha_\mathrm{cool}$ in all three simulations, at the level of $\sim1\%$ or less, we can safely conclude that the data does not present any evidence that the real WD cooling process may be noticeably slower than is prescribed by modern WD cooling models. The best optimised value of the DA WD fraction, $0.77$, is only by a few percent lower than the widely used value of 0.8, and according to the corresponding $S_\mathrm{rel}$, this parameter is moderately constrained. 

For the SFH, we obtained the new and well-constrained parameters describing its overall shape and normalisation, as well as the mean age of the tested star-formation enhancement episode. The mean age of this peak shifted from its old value by \mbox{$\sim0.6$ Gyr} to younger ages and is now \mbox{$\sim2.4$ Gyr}. However, we did not obtain any additional information about its intensity represented by the parameter $\Sigma_\mathrm{p1}$. It is also important to note that the shape-related SFH parameters, $\eta$ and $\zeta$, are strongly correlated. 

All selected IMF parameters were robustly updated. There is a strong negative correlation between its higher-mass slope $\alpha^\mathrm{IMF}_2$ and the overall SFH normalisation $\Sigma_\mathrm{d}$. There are no other SFH-IMF degenerate pairs: the full CMD covers both young and old evolutionary stages as well as massive and low-mass stars, thus providing enough observational constraint for narrowing the allowed regions of the parameter space, such that the SFH-IMF degeneracy becomes less prominent. There is also a visible correlation between the low-mass IMF slope $\alpha^\mathrm{IMF}_0$ and the mass break point corresponding to this interval, $m^\mathrm{IMF}_0$. 

\begin{figure}
\includegraphics[width=1.0\linewidth]{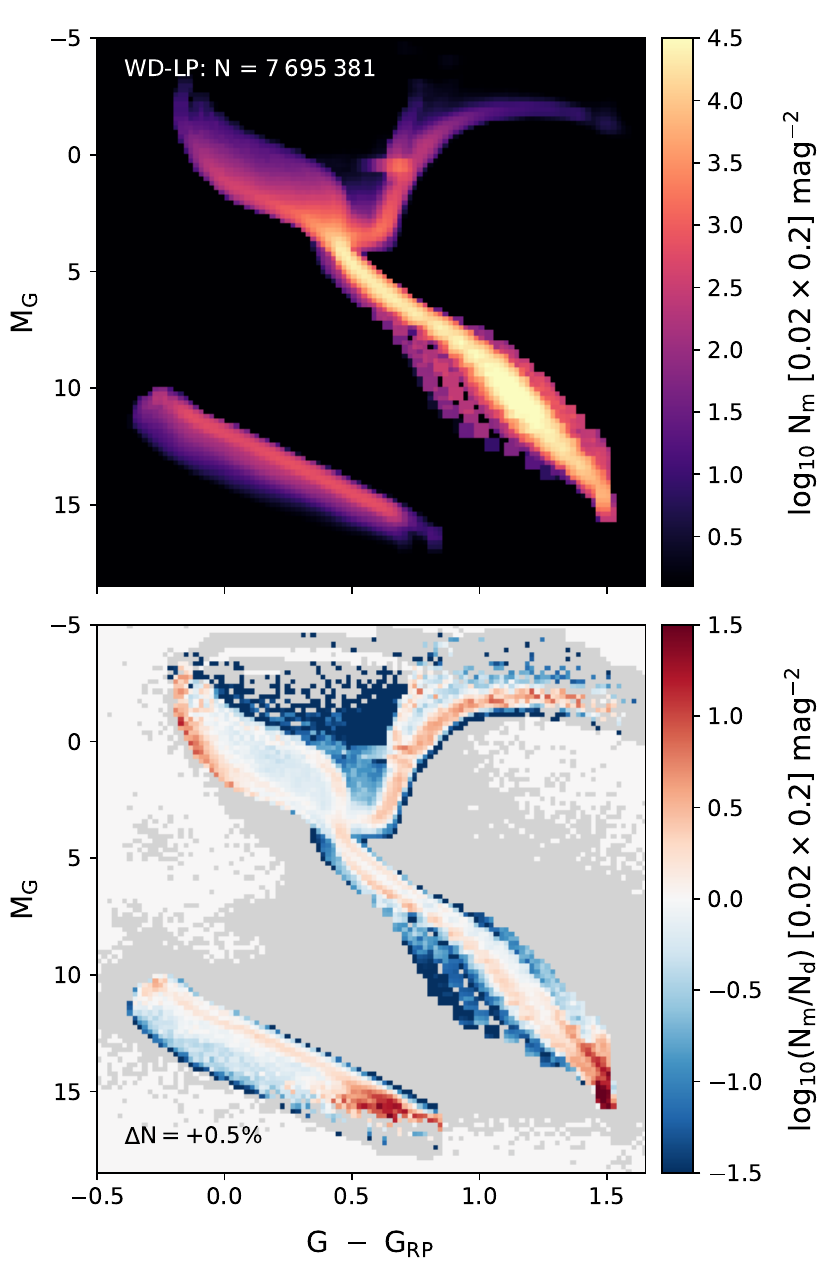}
\caption{Modelled Hess diagram for the calibrated WD-LP model (upper panel) and its comparison to our Gaia sample $S_0$ in terms of logarithmic residuals (bottom panel).}
\label{fig:hess_wdlp}
\end{figure}

After recalibration, the model provides a substantially improved description of all populations, and of the white dwarfs in particular, the WD count residual falls from $+81.4$\% to $+0.5$\% while the good agreement with the giant and main-sequence regions is preserved (\tabref{tab:counts-4models}). 
The Hess diagram predicted by the fiducial WD-LP calibration is shown in \figref{fig:hess_wdlp}. The model reproduces the main sequence, giant branch, lower main sequence, and WD cooling sequence simultaneously, with the largest improvement over the legacy model in the WD region. 

\subsubsection{The new SFH, IMF, and IFMR}
\label{sec:SFR_IMF_IFMR}

\begin{figure*}
\includegraphics[width=\textwidth]{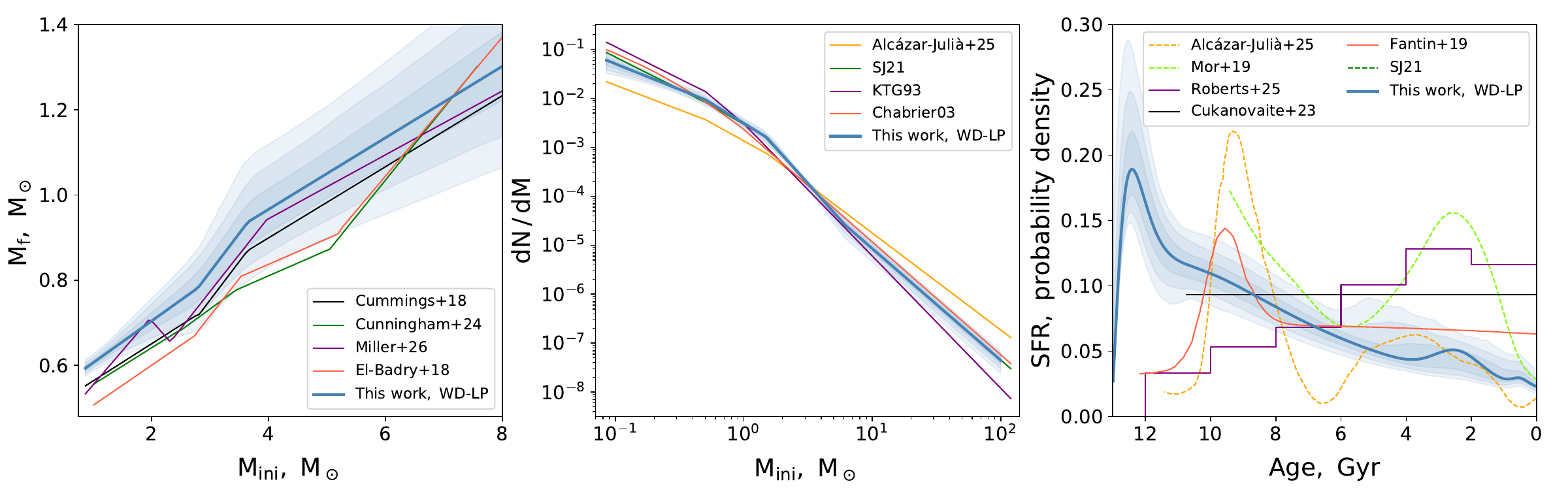}
\caption{Comparison of the functions of the calibrated CALOMERA model to their observational and theoretical determinations in other works. Results of this work are shown in solid blue lines with the shaded regions marking the posterior credible intervals corresponding to 1-, 2-, 3-$\sigma$. \textit{Left:} The inferred initial--final mass relation (IFMR) compared to recent empirical determinations from \citetalias{Cummings2018}, \citet{ElBadry2018}, \citet{Cunningham2024}, and \citet{Miller2026}. \textit{Centre:} The calibrated initial mass function (IMF) compared to \citet[][KTG93]{Kroupa1993_IMF}, \citet{Chabrier2003}, and \citet{AlcazarJulia2025}, as well as our previous calibration in \citetalias{Sysoliatina2021}. \textit{Right:} The recovered star formation history (SFH) of the Solar neighbourhood based on both white dwarfs and non-degenerate stars compared to the literature SFH; results that relied on WDs only are shown in solid lines \citep[][]{Fantin2019,Cukanovaite2023,Roberts2025}, and results that were derived from other populations are plotted with dashed lines \citep[][]{Mor2019,Sysoliatina2021,AlcazarJulia2025}.}
\label{fig:ifmr-imf-sfr}
\end{figure*}

\figref{fig:ifmr-imf-sfr} compares the recovered IFMR, IMF, and SFH with representative literature determinations.

We found the WD population to be especially sensitive to the low-mass regime of the IFMR which sets both the progenitor lifetime and the resulting cooling age, whereas moderate changes in the IMF or SFH mainly rescale the overall normalisation of the WD sequence. 

The inferred IFMR is broadly consistent with its recent empirical determinations \citep[][]{Cummings2018, ElBadry2018, Cunningham2024, Miller2026} (left panel of \figref{fig:ifmr-imf-sfr}). As mentioned in \secref{sec:new_params}, our modelling reliably constrained only the low-mass part of the IFMR represented by $b_0$ and $\alpha^\mathrm{IFMR}_0$ parameters, whose posterior distributions shrink substantially relative to their priors (\tabref{tab:best_param}). The intermediate and high-mass IFMR regimes remain fully prior-dominated. Namely, the MCMC-sampled posterior reproduces the adopted prior; a few additional short simulations showed that this behaviour sustains if we change the IFMR prior. Therefore, we warn the reader of the literal interpretation of the reported uncertainties for the middle and high-mass IFMR intervals. The reported uncertainties are formal, dictated by the assumed prior, but in fact we do not provide any constraints on these IFMR regimes in the framework of our modelling, and thus the differences between our best IFMR and other IFMR forms from the literature should not be overinterpreted. In the present parameterisation and likelihood that relies on star counts in the colour-magnitude space, the WD population strongly constrains the low-mass IFMR normalisation, rather than the detailed shape.

The calibration favours a slightly steeper intermediate-mass IMF than \citetalias{Sysoliatina2021}; because stars in this range dominate the progenitors of present-day WDs, even modest IMF changes propagate into measurable shifts in the WD luminosity function, mass distribution, and number counts.
Despite this change, the calibrated IMF remains close to its commonly adopted Galactic forms across the full mass range (middle panel of \figref{fig:ifmr-imf-sfr}). Above $\sim1\,M_\odot$, the inferred slope is similar to the canonical \citet{Kroupa1993_IMF} IMF, while at the highest masses it approaches the \citet{Chabrier2003} form. Below $1\,M_\odot$, the IMF exhibits a modest flattening relative to a single power law, remaining consistent with the turnover expected from MW field-star IMFs.

The star formation history recovered from the joint fit is shown in the right panel of \figref{fig:ifmr-imf-sfr}. It is characterised by an early peak associated with the thick-disc phase followed by a power-law decline toward the present, the thin disc phase. The new thin-disc SFH is characterised by the shape parameters $\zeta = 0.58^{+0.20}_{-0.20}$ and $\eta = 5.02^{+0.99}_{-0.98}$ (\tabref{tab:best_param}), together with two secondary episodes of enhanced star formation: one at \mbox{$\tau_\mathrm{p1} = 2.42^{+1.04}_{-0.78}$ Gyr} (optimised in this study) and another at \mbox{$\tau_\mathrm{p2} = 0.5$ Gyr} (adopted from \citetalias{Sysoliatina2021}). We emphasise that only the age of this first episode is constrained by the data; its amplitude $\Sigma_\mathrm{p1}$ returns the prior (\tabref{tab:best_param}). The recovered local star formation history of the Milky Way is thus best described as an early peaked function with a subsequent  declining continuum punctuated by two moderate recent bursts.

This declining SFH shape places our joint fit closer to the CMD-based reconstructions of the Solar neighbourhood (dashed curves in the SFH panel of \figref{fig:ifmr-imf-sfr}; \citealt{Mor2019, AlcazarJulia2025}) than to the near-constant or even rising histories obtained from several WD-only inversions (solid curves; \citealt{Fantin2019, Cukanovaite2023}). Because the priors for $\zeta$ and $\eta$ were deliberately made wide enough to admit constant ($\zeta = \eta/2$) and even rising solutions, the recovery of a declining SFH shape is a genuine outcome of the fit rather than a consequence of the prior. We note that, unlike the WD-only reconstructions, our SFH is required to reproduce the main sequence, giants, LMS, and WD sequence simultaneously under a single IMF and chemical model in the form of age-metallicity relation, and is therefore not subject to the WD-specific modelling freedom that may contribute to the diversity of published WD-only histories.

\subsubsection{White dwarf masses, temperatures, and luminosity function}

\begin{figure}
    \centering
    \includegraphics[width=1\linewidth]{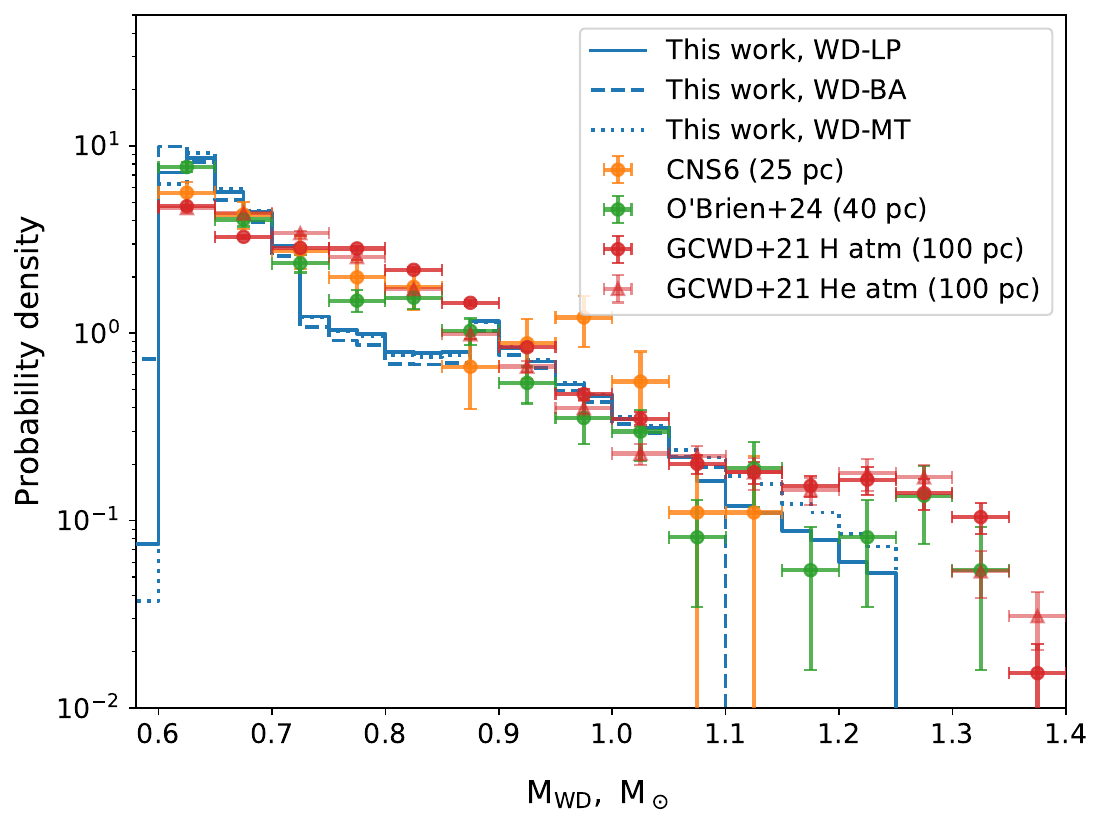}
    \includegraphics[width=1\linewidth]{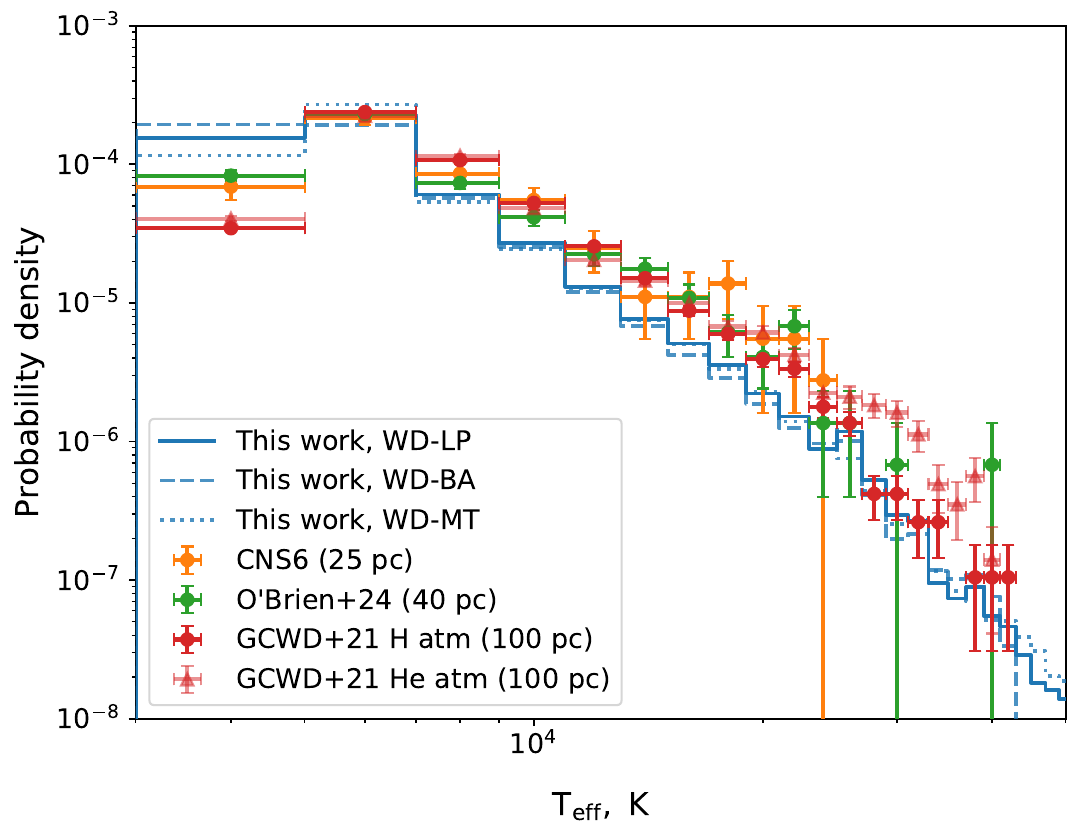}
    \caption{\textit{Top:} WD mass distributions predicted by the calibrated model, evaluated within a 100~pc sphere, for the three cooling prescriptions, compared with local observational samples. \textit{Bottom:} corresponding effective-temperature distributions. The comparisons are shown for the 25 pc CNS sample, the 40 pc sample of \citet{OBrien2024}, and the 100 pc GCWD21 sample.  The GCWD21 results are shown separately for hydrogen- and helium-atmosphere WDs.}
    \label{fig:wd_mass}
\end{figure}

We compare the calibrated model with the observed WD mass and effective-temperature distributions of local samples based on different selection criteria: the 25-pc CNS5 sample, the 40-pc sample of \citet{OBrien2024}, and the 100-pc GCWD21 sample (\figref{fig:wd_mass}). The model distributions are shown for a 100~pc realisation; the corresponding 25 and 40~pc realisations are nearly identical when normalised. For consistency, we restrict the comparison to $M_\mathrm{WD} > 0.6\,M_\odot$, since the calibrated IFMR implies a minimum WD mass near $0.6\,M_\odot$ and lower-mass WDs are expected to form predominantly through binary evolution, which we do not model. The same mass cut is applied to the observational samples. Although this mass cut truncates the WD mass distribution and modifies the shapes of the effective-temperature and cooling-age distributions, it ensures a valid comparison between the data and model, both affected by this mass-related incompleteness effect in the same way. For the two smaller samples of 25 and 40 pc, the model reproduces the effective-temperature and WD mass distributions. The three cooling prescriptions yield similar distributions, confirming that these observables are set by the inferred IFMR, SFH, and IMF rather than by the cooling model.

The comparison with GCWD21 requires caution. As discussed by \citet{OBrien2024}, the atmospheric-parameter determinations systematically underestimate the masses and effective temperatures of cool DA-type WDs. This is partly associated with incomplete cool-atmosphere opacities, particularly in the Ly$\alpha$ red wing, bias photometric masses and effective temperatures \citep{ McCleery2020, OBrien2024}. The updated opacities of \citet{Sahu2025} reduce this effect, but do not eliminate this discrepancy; these corrections are not currently applied to the GCWD21 sample. In addition, \citet{Golovin2024} find a non-negligible fraction of spurious candidates at the faint end (${\rm M_G} \gtrsim 16$~mag). Since neither effect is corrected here, the volume-complete 25- and 40-pc samples provide more robust benchmarks for physical properties, whereas GCWD21 remains useful for number counts down to the onset of incompleteness at the faint end (${\rm M_G} \gtrsim 16$~mag), primarily associated with low parallax signal-to-noise ratios and large harmonic-amplitude values in \textit{Gaia}.

The white dwarf luminosity function (WDLF) is one of the strongest joint constraints on the SFH and the cooling models, since it combines the progenitor population, the IFMR, the SFH, and the cooling times. Defined as a number density of white dwarfs per unit luminosity or absolute magnitude interval, WDLF is independent of the studied volume assuming that the population is homogeneous. In practice, however, WDLF estimation is often based on magnitude-limited or incomplete data. Additionally, the WD population is predominantly old and therefore dynamically heated. As a result, old white dwarfs are increasingly underrepresented in smaller volumes.

\begin{figure}
    \centering
    \includegraphics[width=1\linewidth]{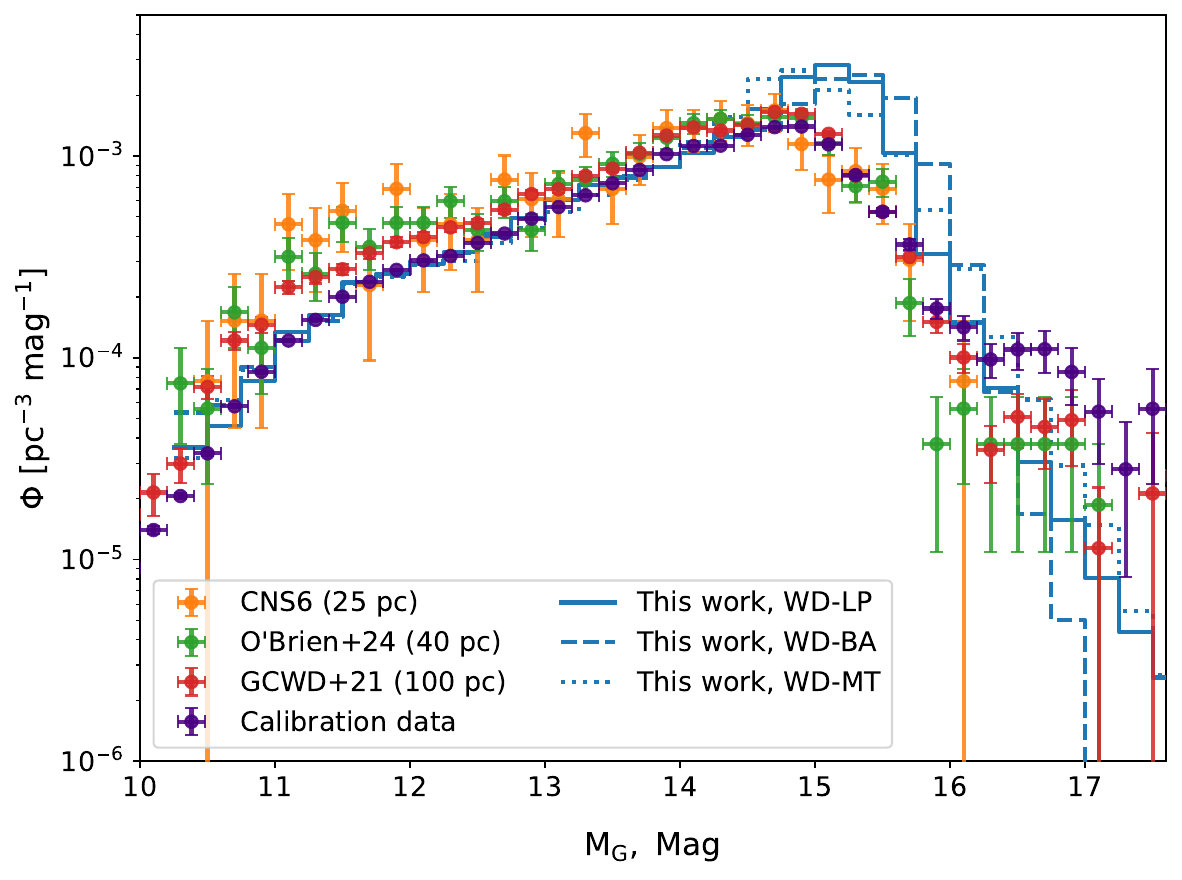}
    \caption{White dwarf luminosity functions predicted by the calibrated models within a 100~pc sphere using the three WD cooling sequences. The blue histograms show the corresponding 100~pc model predictions for the three cooling libraries, while the data points represent the different observational samples. The purple points indicate the WDLF of the calibration sample.}
    \label{fig:wdlf_model_data}
\end{figure}

Thus, the data-based WDLF remains sensitive to the sampled volume, as illustrated in \figref{fig:wdlf_model_data} for the 25-, 40-, and 100-pc samples, together with the calibration sample \secref{sec:main_sample}. We constructed the observed WDLFs using fixed magnitude bins, with uncertainties derived from Poisson statistics. For comparison, the model predictions are evaluated within a 100-pc sphere using each of the three calibrated WD cooling libraries. The 25- and 40-pc WDLFs agree reasonably well with the model at brighter magnitudes, but their faint ends are dominated by low-number statistics. The larger 100-pc sample provides better statistics over the full magnitude range considered here.

Over most of the luminosity range, the calibrated models reproduce the WDLF well, with some differences between the cooling implementations. The cooling prescriptions begin to diverge towards the faint end: WD-BA shows a relatively sharp cut-off, WD-MT and WD-LP declines more gradually. Since the predicted counts are small in this regime, these differences should be interpreted with caution and do not provide a robust basis for distinguishing between the cooling prescriptions.

The predicted WDLF broadly agrees well with the 100-pc GCWD21 sample, although differences remain near the peak and at the faint end. Although \citet{Golovin2024} suggest that the faint-end counts may be slightly overestimated, this would primarily affect the slope of the terminal drop-off and is unlikely to substantially alter the position or amplitude of the WDLF peak. Nevertheless, additional systematics may remain, including residual incompleteness among the coolest WDs, uncertainties in the cooling models and inferred SFH, and observational uncertainties, which cannot be fully resolved with the present sample.

\begin{figure}
    \centering
    \includegraphics[width=1\linewidth]{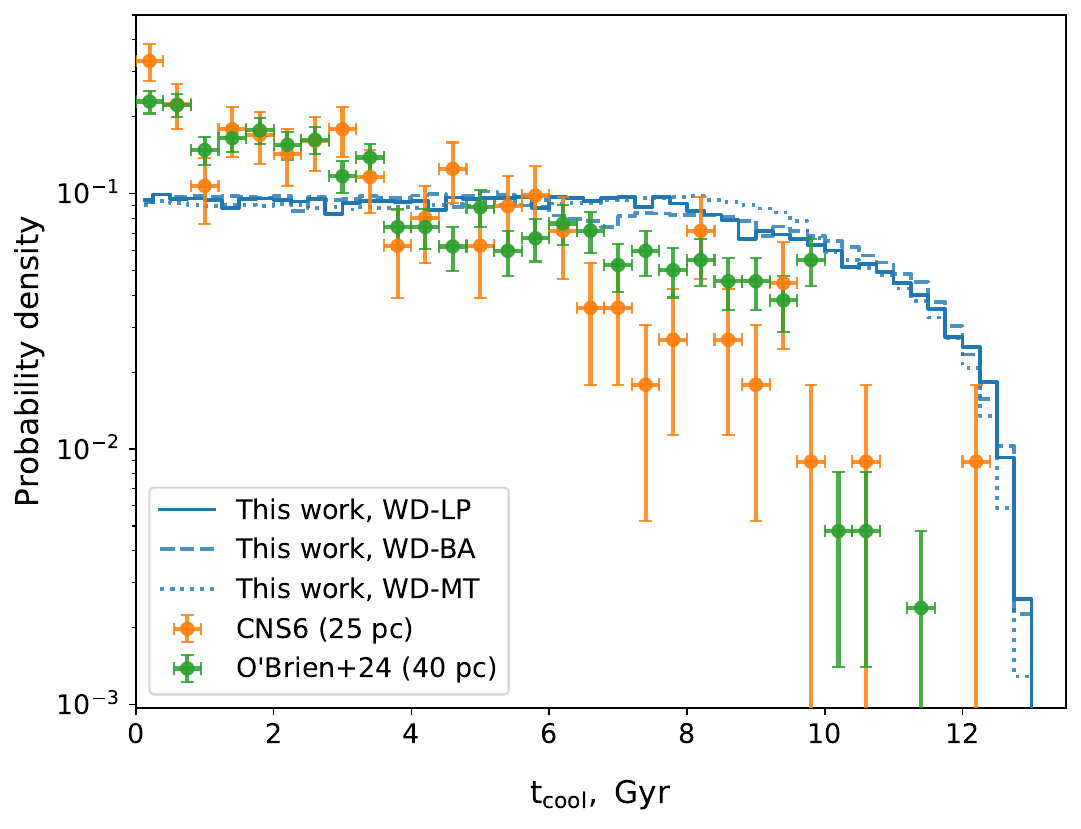}
    \caption{Comparison of the predicted white dwarf cooling-age distributions with the 25 pc and 40 pc volume-complete samples. The model curves correspond to a 40~pc realisation.}
    \label{fig:wd_coolingage}
\end{figure}

The inferred cooling-age distributions are shown in \figref{fig:wd_coolingage} for the volume-complete 25- and 40-pc samples. Because the cooling age is derived independently of the progenitor lifetime, this comparison probes the cooling sequences rather than the SFH. The observed cooling-age distributions exhibit a steep decline with increasing cooling age in both samples, whereas the model predicts a flatter distribution that declines more gradually towards the oldest cooling ages. The models underpredict the youngest WDs and predict relatively more WDs at intermediate and old cooling ages.
The three libraries diverge only at the intermediate cooling ages, \mbox{$t_\mathrm{cool} \sim 6-8$ Gyr}, where crystallisation-related delays accumulate, but the current sample sizes do not distinguish between them.

\begin{figure}
    \centering
    \includegraphics[width=1\linewidth]{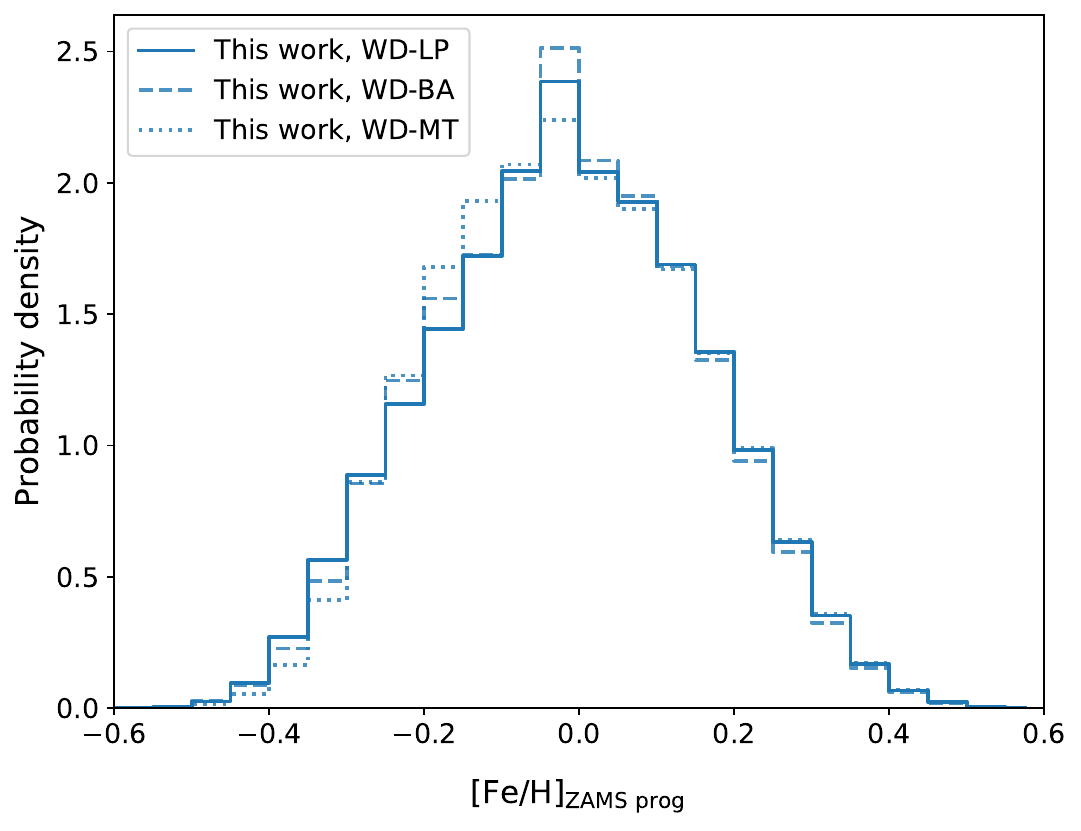}
\caption{Predicted initial metallicity distribution of white dwarf progenitors within the local 100 pc sample. The progenitor metallicities are inferred from the calibrated age--metallicity relation combined with the metallicity-dependent main-sequence lifetimes.}
    \label{fig:met_WD_proh}
\end{figure}

A further advantage of the joint synthesis is that it links each present-day WD to its progenitor. Combining the calibrated IFMR with the metallicity-dependent progenitor lifetimes and the AMR allows us to infer the initial metallicity of every WD progenitor. The resulting metallicity distribution within 100 pc is shown in \figref{fig:met_WD_proh}. It peaks at solar metallicity, with the majority of progenitors spanning approximately $-0.4 \lesssim \mathrm{[Fe/H]} \lesssim 0.4$. This distribution provides a direct connection between the present-day WD population and the chemical evolution of the Galactic disc.

\begin{figure}
    \centering
    \includegraphics[width=1\linewidth]{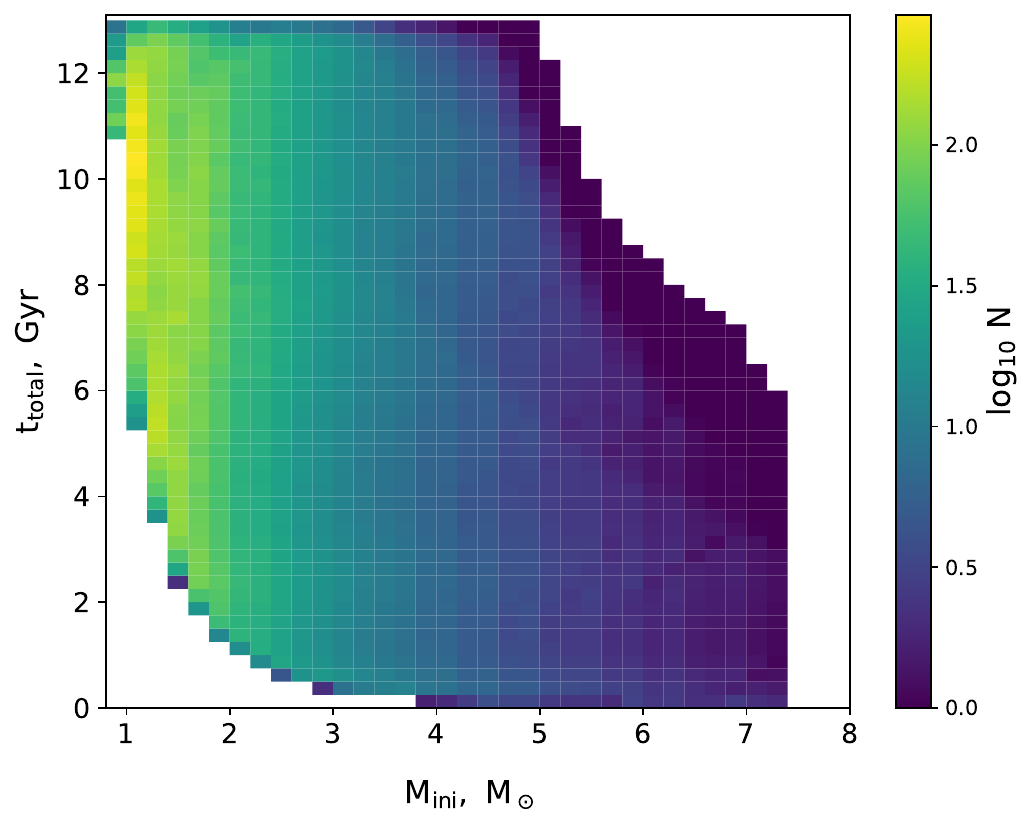}
    \caption{Distribution of white dwarf progenitor initial mass, $M_{\rm ini}$, and total age, for the calibrated WD-LP model. The colour scale shows the logarithm of the number of objects in each mass--age bin.}    
    \label{fig:Mini_age}
\end{figure}

The corresponding progenitor initial mass--total age distribution for the calibrated WD-LP model within a 100~pc sphere is shown in \figref{fig:Mini_age}. As expected, the population is dominated by low-mass progenitors, reflecting both the declining IMF and the longer main-sequence lifetimes of low-mass stars, while the lower envelope is set by the minimum lifetime required for a star of a given initial mass to evolve into a WD.

\subsection{Predicted Galactic white dwarf distribution}
\label{sec:PredictionOfModel}
Using the fiducial WD-LP calibration, we generate predictions for the spatial distribution and observability of WDs across the Milky Way disc. These use the radially extended version of the model \citep{Sysoliatina2022}, in which the disc is represented by 91 concentric annuli spanning \mbox{$4.5 \leq R \leq 13.5$ kpc} at \mbox{$\Delta R = 0.1$ kpc} spacing. At each radius the vertical structure (\equref{eq:rho_z}) is solved self-consistently for the thin disc, thick disc, gas, stellar halo, and dark matter halo. The IMF, IFMR, cooling prescription, and DA/DB fraction are held independent of radius, while the thin-disc surface density, SFH shape, and AMR vary with $R$ to reproduce the inside-out growth and chemical evolution of the disc \citep{Sysoliatina2022}. Each annulus is treated independently, with no radial migration between annuli and assuming that the radial and vertical motions are decoupled.

\begin{figure*}
    \centering
    \includegraphics[width=1\linewidth]{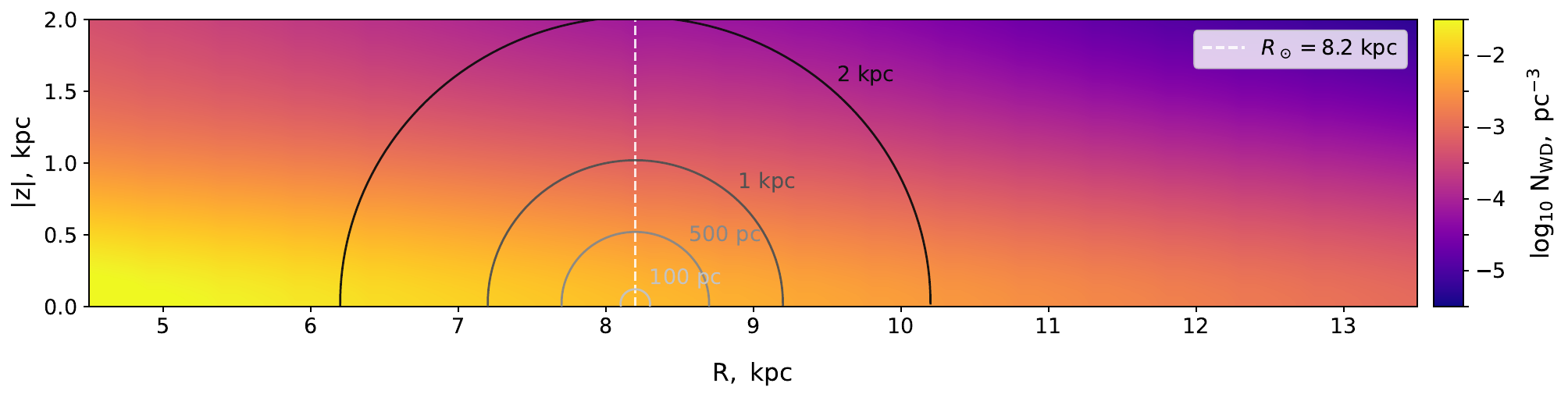}
    \includegraphics[width=1\linewidth]{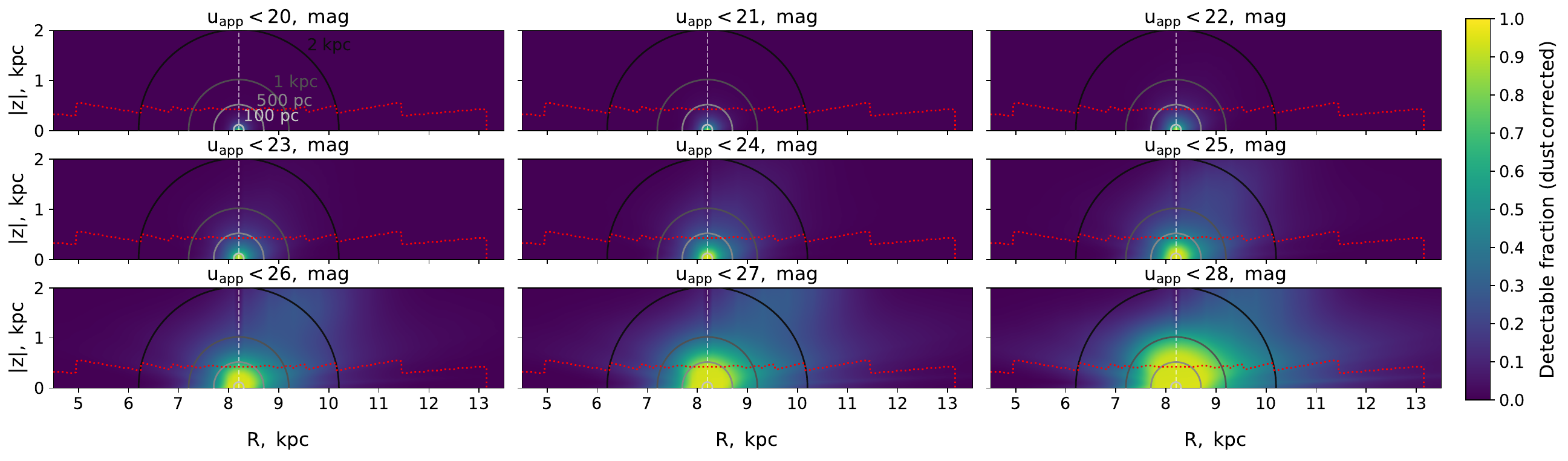}
    \includegraphics[width=1\linewidth]{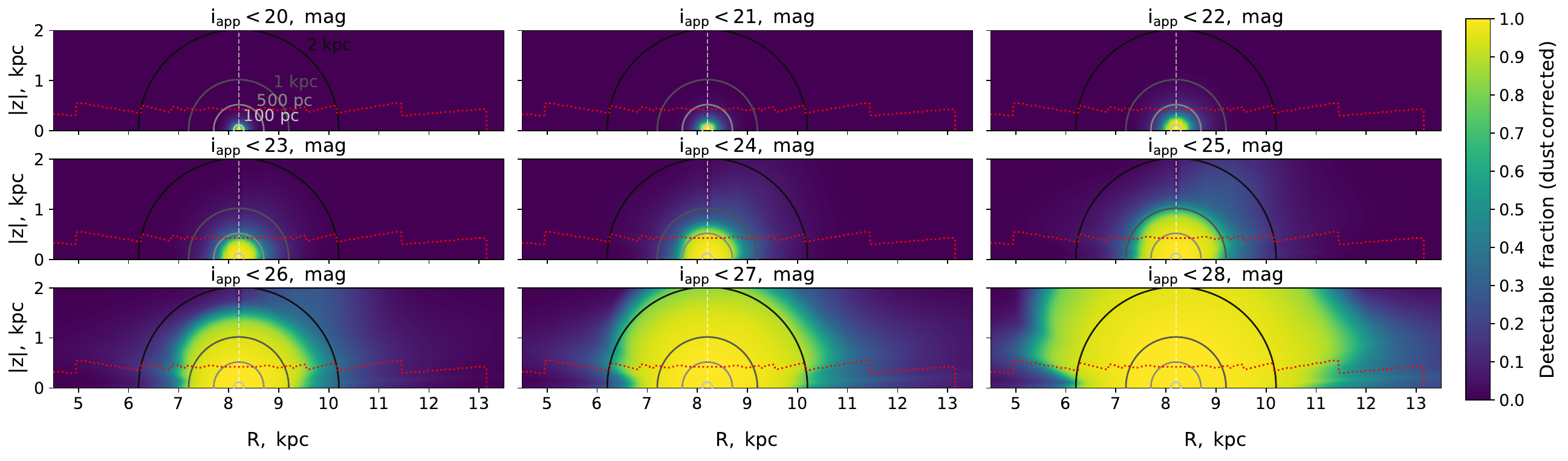}
    \caption{Realisation based on the WD-LP calibration. \textit{Top:} predicted WD number-density map in the $(R,z)$ plane. The dashed vertical line marks $R_\odot$, and the circular contours show heliocentric distances of 100 pc, 500 pc, 1 kpc, and 2 kpc. \textit{Middle and bottom:} predicted observable fraction of WDs in the SDSS $u$ and $i$ bands, respectively, including extinction from the 3D dust map of \citet{Lallement2022} with red dotted line showing its applicable range.}

    \label{fig:WD_Rz_map}
\end{figure*}

Evaluating the density field over the full $(R,z)$ grid, to \mbox{$|z| = 2$~kpc} at \mbox{$\Delta z = 2$~pc} resolution, yields the WD density map in the upper panel of \figref{fig:WD_Rz_map}. The density decreases with both radius and height, reflecting the declining stellar surface density, the radial variation of the SFH and AMR, the self-consistent potential, and the age-dependent dynamical heating. Integrating over heliocentric spheres gives approximately $2.1\times10^{6}$, $12.3\times10^{6}$, and $58.6\times10^{6}$ WDs within 500~pc, 1~kpc, and 2~kpc respectively (\tabref{tab:wd_numbers_spheres}). Within 100~pc, the predicted count agrees with the $\rm 1/V_{\max}$ corrected GCWD21 count to within approximately 10\%.

\begin{table}
    \centering
    \small
    \setlength{\tabcolsep}{3pt}
    \begin{tabular}{lcccc}
        \toprule
        Volume & 100 pc & 500 pc & 1 kpc & 2 kpc \\
        \midrule
        WDs &
        $1.9 \times 10^{4}$ &
        $2.1 \times 10^{6}$ &
        $12.3 \times 10^{6}$ &
        $58.6 \times 10^{6}$ \\
        \bottomrule
    \end{tabular}
    \caption{Predicted total number of WDs within spherical volumes of different radii in the calibrated model.}
    \label{tab:wd_numbers_spheres}
\end{table}

Beyond the intrinsic density, the model predicts the fraction of the WD population recoverable by surveys of different depth. For each $(R,z)$ cell we construct the local WDLF and compute the cumulative fraction of objects brighter than a limiting apparent magnitude, separately per passband. We show the SDSS $u$ and $i$ bands including extinction from the three-dimensional dust map of \citet{Lallement2022}\footnote{We use the $10\times10\times0.8$ kpc dust map with the extinction coefficients of \citet{Schlafly2011, Casagrande2018}, evaluated in the meridional ($\phi = 0$) plane by mapping each $(R,z)$ cell to its heliocentric $(l, b, d)$ coordinates.}, with the line of sight defined toward $l = 0^\circ$ for $R \leq R_\odot$ and $l = 180^\circ$ for $R > R_\odot$ and $b$ set by the pixel height. Reddening is interpolated along this sightline, averaged over the nearest HEALPix directions with inverse-angular-distance weights, and held fixed beyond the map's range.

The lower panels of \figref{fig:WD_Rz_map} show the resulting observable fraction for several limiting magnitudes: as depth increases, progressively older and fainter WDs become accessible to larger distances and heights. 

The detection fractions within heliocentric spheres are summarised in \tabref{tab:wd_completeness_allbands}, for the \textit{Gaia} $G$ band with and without extinction and for the dust-corrected SDSS $u$ and $i$ bands. The detectable fraction rises with survey depth and falls with distance. Dust has only a minor effect within 100 pc but reduces the observable fraction at larger distances, most strongly in the blue $u$ band and weakly in the red $i$ band. In the \textit{Gaia} $G$ band, a survey complete to $G \simeq 26$ recovers about 95\% of WDs within 1 kpc, while a comparably complete census within 2 kpc requires $G \simeq 28$ ($\sim$97\%).

\setlength{\extrarowheight}{0.25em}
\begin{table*}
\centering
\small
\setlength{\tabcolsep}{4pt}
\caption{Predicted observable fraction (\%) of the WD population within spherical volumes centred on the Sun as a function of limiting magnitude. The Gaia $G$-band predictions are shown both without and with extinction, while the SDSS $u$ and $i$ bands include extinction from the three-dimensional dust map of \citet{Lallement2022}.}
\label{tab:wd_completeness_allbands}
\begin{tabular}{c|cccc|cccc|cccc|cccc}
\toprule
&
\multicolumn{4}{c|}{Gaia $G$ (no dust)} &
\multicolumn{4}{c|}{Gaia $G$ (dust)} &
\multicolumn{4}{c|}{SDSS $u$ (dust)} &
\multicolumn{4}{c}{SDSS $i$ (dust)}\\

${\rm Mag}_{\rm lim}$ &
100 pc & 500 pc & 1 kpc & 2 kpc &
100 pc & 500 pc & 1 kpc & 2 kpc &
100 pc & 500 pc & 1 kpc & 2 kpc &
100 pc & 500 pc & 1 kpc & 2 kpc\\
\midrule
20 & 78.8 & 6.6 & 1.1 & 0.1 & 77.9 & 5.9 & 0.9 & 0.1 & 32.9 & 3.7 & 0.6 & 0.1 & 87.7 & 5.0 & 0.6 & 0.1\\
21 & 95.8 & 15.7 & 3.9 & 0.5 & 95.7 & 14.2 & 3.2 & 0.4 & 46.8 & 8.2 & 2.1 & 0.3 & 97.2 & 14.2 & 2.5 & 0.3\\
22 & 98.2 & 32.5 & 10.4 & 2.1 & 98.2 & 29.7 & 8.9 & 1.6 & 65.4 & 15.3 & 5.2 & 1.1 & 99.1 & 33.1 & 8.3 & 1.2\\
23 & 99.7 & 60.4 & 22.9 & 6.5 & 99.6 & 55.2 & 19.9 & 5.3 & 84.9 & 24.9 & 10.7 & 3.2 & 100.0 & 65.0 & 21.3 & 4.5\\
24 & 100.0 & 92.1 & 44.5 & 15.5 & 100.0 & 88.9 & 39.5 & 13.2 & 92.7 & 36.5 & 18.9 & 7.4 & 100.0 & 94.4 & 45.9 & 13.3\\
25 & 100.0 & 97.3 & 78.8 & 32.3 & 100.0 & 96.9 & 71.4 & 28.0 & 92.7 & 51.3 & 29.3 & 14.1 & 100.0 & 98.1 & 83.6 & 31.5\\
26 & 100.0 & 99.0 & 95.8 & 60.0 & 100.0 & 98.8 & 94.8 & 52.9 & 92.7 & 70.6 & 41.9 & 23.4 & 100.0 & 99.6 & 96.8 & 62.9\\
27 & 100.0 & 100.0 & 98.2 & 91.9 & 100.0 & 99.9 & 97.9 & 87.5 & 92.7 & 88.1 & 58.9 & 34.6 & 100.0 & 100.0 & 98.9 & 94.0\\
28 & 100.0 & 100.0 & 99.7 & 97.3 & 100.0 & 100.0 & 99.4 & 96.7 & 92.7 & 92.7 & 79.1 & 49.0 & 100.0 & 100.0 & 99.9 & 98.0\\
\bottomrule
\end{tabular}
\end{table*}

\begin{figure}
    \centering
    \includegraphics[width=\linewidth]{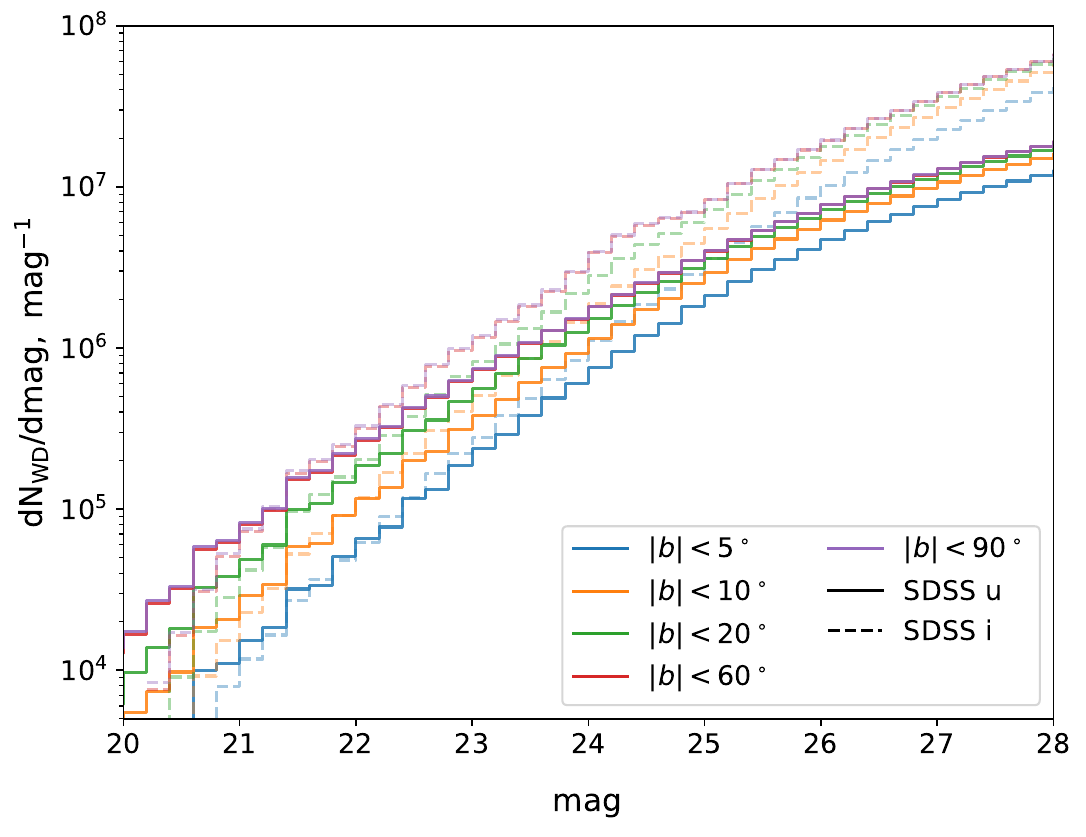}
    \caption{Predicted differential WD number counts in the $u$ (solid line) and $i$ (faint dotted line) SDSS bands as a function of apparent magnitude for different Galactic latitude cuts using the WD-LP calibration. The predictions include extinction from the adopted three-dimensional dust map \citep{Lallement2022} and are computed from the calibrated model by integrating the WD density distribution over the corresponding survey volume.}
    \label{fig:dNdG_WD}
\end{figure}

The differential number counts in \figref{fig:dNdG_WD} are generated from the same forward model and give the predicted number of WDs per 0.1~mag interval. The sky is divided into longitude bins and heliocentric distance shells mapped onto Galactocentric radius and height. For each latitude cut, the self-consistent density profile is integrated over the volume, and model absolute magnitudes are converted to apparent magnitudes using the distance modulus and the dust extinction above. In both the $u$ and $i$ bands the counts rise steeply toward fainter magnitudes as older, fainter WDs become detectable, and broader latitude cuts contain more WDs at all magnitudes by sampling larger volumes and greater heights.

\subsection{Extension to the substellar regime}
\label{sec:results_bd}
Because the calibrated framework maps ages and masses onto observables for any evolutionary sequence, it extends naturally to the substellar regime, whose present-day population is sensitive to the low-mass IMF and the SFH. We generate synthetic photometry and physical properties for brown dwarfs from the chemical-equilibrium ATMO\footnote{\url{https://www.erc-atmo.eu/}} models \citep{Phillips2020_ATMO}, applying the WD-LP calibration without any additional tuning, by interpolating the model masses and ages from the grids in the same way as for the WDs.

We stress that the substellar population was not included in the calibration and that the predicted number densities are particularly sensitive to the low-mass IMF slope and the SFH; the results below are therefore predictions of the Galactic model rather than independent constraints on the substellar population. As an initial check, we compare the predicted population within 25 pc with the CNS5 census \citep{Golovin2023}, which is essentially complete to about L8, adopting the stellar--substellar boundary at $0.076\,M_\odot$ ($\approx$ L1; \citealt{Dupuy2012, Pecaut2013})\footnote{Updates available at \url{https://www.pas.rochester.edu/~emamajek/EEM_dwarf_UBVIJHK_colors_Teff.txt}} and assigning spectral types through the predicted effective temperatures.
The CNS5 catalogue is expected to be essentially complete for L dwarfs within this volume, whereas its T-dwarf census is incomplete, and only a small number of Y dwarfs are known. Compared to the observed L-dwarf sample, CALOMERA predicts approximately 20\% more L dwarfs than are catalogued. Assuming the completeness estimate for the L-dwarf sample is accurate, this discrepancy most likely reflects an overprediction of the local L-dwarf population rather than incompleteness in the observations. It may instead indicate residual uncertainties in the low-mass IMF, the adopted SFH, or the brown-dwarf evolutionary models, none of which is directly constrained by the WD-LP calibration. This discrepancy may also arise from sample incompleteness or from statistical fluctuations associated with the relatively small 25~pc volume.
For the T-dwarf regime, the model predicts roughly 70\% more objects than catalogued within 25 pc, consistent with a large undiscovered T-dwarf population, and it predicts a Y-dwarf population that is almost entirely absent from CNS5. Taken together, the catalogued L, T, and Y dwarfs represent only about 20\% of the total L, T, and Y population the model predicts within 25~pc (\tabref{tab:bd_numbers_spheres}), the remainder being dominated by the coolest, as-yet-undiscovered faint T and Y dwarfs, and possibly more L dwarfs. 

\begin{table}
\centering
\small
\setlength{\tabcolsep}{4pt}
\caption{Predicted numbers of L, T, and Y dwarfs within heliocentric spheres for the calibrated WD-LP model.}
\label{tab:bd_numbers_spheres}
\begin{tabular}{lcccc}
\toprule
Volume & 25~pc & 100~pc & 500~pc & 1~kpc \\
\midrule
Brown dwarfs &
$3.70\times10^{3}$ &
$1.54\times10^{5}$ &
$1.23\times10^{7}$ &
$6.82\times10^{7}$ \\
\bottomrule
\end{tabular}
\end{table}

Extending the predictions to larger volumes yields the L-, T-, and Y-dwarf counts listed in \tabref{tab:bd_numbers_spheres}. Within a sphere of radius 100\,pc centred on the Sun, we predict approximately $10^5$ substellar objects. As a practical application, we estimate the foreground L, T, and Y dwarf population in deep \textit{JWST} extragalactic fields using NIRCam F200W photometry from the ATMO NIRCam grids, for the JADES \citep{Eisenstein2026}, CEERS \citep{Finkelstein2025}, and COSMOS-Web \citep{Casey2023} geometries (\tabref{tab:JWST_BD_counts}, full details in \appref{app:BDPredictions}). The foreground is dominated by L dwarfs at bright magnitudes and by T and then Y dwarfs at fainter limits. This is relevant because Galactic brown dwarfs have been recognised as contaminants of compact red sources in deep \textit{JWST} imaging, some photometrically selected Little Red Dot candidates \citep[e.g.,][]{Matthee2024} are in fact Galactic T and Y dwarfs \citep{Langeroodi2023}, and further candidates have been found in JADES, CEERS, and RUBIES \citep{Hainline2024, Tu2025}. Our model predicts a substantially larger foreground population than is currently identified. In particular, approximately 1306 brown dwarfs are expected to be detectable to a limiting magnitude of $F200W < 30$\,mag across the combined JADES, CEERS, and COSMOS-Web survey areas (\appref{app:BDPredictions}; \tabref{tab:JWST_BD_counts}). Relative to a limiting magnitude of $F200W < 28$\,mag, this corresponds to an additional $\sim$24, 14, and 465 brown dwarfs in the JADES, CEERS, and COSMOS-Web fields, respectively, detectable in the interval $28 < F200W \lesssim 30$\,mag. While only a fraction of these objects are expected to satisfy the colour and morphology criteria used to select compact high-redshift sources, this result suggests that the potential level of brown-dwarf contamination in deep \textit{JWST} surveys may be significantly larger than implied by the currently confirmed contaminants.
Predictions for the MKO $J$ detectability of the substellar population are given
in \appref{app:BDPredictions}.

\section{Discussion}
\label{Sec:Discussion}

\subsection{Joint constraints from white dwarfs and non-degenerate stars}

This work demonstrates that white dwarfs provide constraints on Galactic population synthesis that are complementary to those from non-degenerate stars. The WD cooling sequence is sensitive to the cumulative SFH, the IFMR, the IMF, and the adopted cooling physics, whereas the main sequence and giant branch constrain the stellar content, the IMF, and the vertical Galactic structure. Previous SFH reconstructions based on WDs or on non-degenerate stars have often yielded qualitatively different results, and the two populations have rarely been fitted simultaneously within a single self-consistent Galactic model. By modelling WDs together with the main sequence and giant branch, \textsc{CALOMERA} framework provides a joint calibration in which the same SFH, IMF, AMR, and spatial structure are required to reproduce the full colour--magnitude diagram.

Because a single SFH must reproduce the entire CMD rather than the WD sequence alone, the recovered history is not subject to the WD-specific modelling freedom that may contribute to the diversity of published WD-only reconstructions. As described in \secref{sec:SFR_IMF_IFMR}, the joint fit favours a declining SFH, placing it closer to CMD-based determinations of the Solar neighbourhood \citep{Mor2019, AlcazarJulia2025} than to the near-constant or rising histories obtained from several WD-only histories \citep{Fantin2019, Cukanovaite2023}. This should be interpreted as the present-day local stellar age distribution implied by the adopted Galactic model, rather than as the in-situ SFH at $R_\odot$ (see \secref{sec:migration_caveat}). Nevertheless, the inferred early SFH peak is not independently established by the local WD samples, whose cooling-age distributions show some tension with the model predictions (\secref{sec:calibrated_model}), and its amplitude remains conditional on the adopted stellar-evolution and WD cooling prescriptions.


\subsection{The faint-end of the CMD}
\label{sec:faint_end}

The most significant remaining discrepancy occurs at the faint end of the CMD ($M_\mathrm{G} \gtrsim 15$\,mag), where systematic residuals are present in both the lower main sequence and the coolest WDs. Although the LMS and WD populations occupy distinct regions of the CMD and are described by independent stellar-evolution and cooling models, both residuals occur close to the faint observational limit of the survey. For the main calibration sample, these may partly reflect declining completeness or measurement quality near the survey limit. However, the nearby 25- and 40-pc WD samples are less affected by the \textit{Gaia} magnitude limit, and their discrepancies cannot be explained by this effect alone. Residual incompleteness may nevertheless persist, particularly for unresolved binaries that are difficult to identify photometrically \citep{OBrien2024}. Other possible contributions include uncertainties in the low-mass IMF, the inferred SFH, the adopted Galactic structure, and the stellar-evolution and WD cooling models.

A possible common contribution is the deterioration of astrometric and photometric quality towards the faint limit. Although \textit{Gaia} DR3 is expected to be essentially complete to $G \simeq 20.5$ over most of the sky \citep{CantatGaudin2023}, the completeness and measurement quality of the faintest sources may nevertheless decline within the magnitude range relevant to our sample. However, the present analysis does not isolate this effect from other possible sources of the discrepancy. These include residual deficiencies in the adopted stellar-evolution or WD cooling models, uncertainties in the low-mass IMF, the treatment of unresolved binaries, atmospheric models and synthetic photometry, as well as catalogue selection effects. Disentangling these contributions will require a more detailed assessment of both the observational selection function and the model uncertainties, which is beyond the scope of the present work \citep[e.g.,][]{CantatGaudin2023}.

\subsection{White dwarf cooling models}

The three adopted cooling libraries yield broadly consistent fits after calibration (\tabref{tab:counts-4models}), indicating that the recovered Galactic parameters do not depend strongly on the choice of modern cooling sequence. The residual model-to-model differences are confined primarily to the oldest and faintest WDs and plausibly arise from differing treatments of crystallisation, phase separation, envelope structure, electron-conduction opacities, and atmospheric boundary conditions \citep{Althaus2010, Salaris2010, Blouin2020, Camisassa2021, Sahu2025}. The sensitivity to the adopted cooling prescription depends on the observable: differences between the three cooling model realisations are present in both the WD mass and effective-temperature distributions, particularly at lower masses and temperatures, but become more pronounced towards the faint end of the WDLF. The calibration adjusts the SFH and IFMR such that all three cooling libraries reproduce the observed CMD comparably well, illustrating the degeneracy between Galactic evolution parameters and the details of WD cooling physics. 

An additional uncertainty arises from the treatment of progenitor metallicity. Although metallicity-dependent main-sequence lifetimes are included, the adopted WD cooling sequences assume solar-metallicity progenitors, neglecting the effects of progenitor metallicity on WD core composition, envelope structure, and subsequent cooling. Together with the limitations of the global cooling-delay parameter, this may contribute to the remaining discrepancies between the inferred SFH and the observed local WD distributions. A fully metallicity-dependent treatment is currently limited by the available cooling grids.

Furthermore, the adopted DA/DB cooling prescriptions do not reproduce the additional cooling delays associated with the anomalously delayed Q-branch population \citep{Cheng2019}, while such WDs and other spectral types are not explicitly excluded from the observational samples. Their potentially different cooling histories may therefore contribute to the discrepancies in the observed cooling-age distributions.

Future larger and deeper WD samples, together with independent stellar-age constraints and improved cooling models, may help disentangle these uncertainties.

\subsection{The initial--final mass relation}

The calibrated IFMR is one of the most important ingredients controlling the WD population, since its low-mass end sets the progenitor lifetimes and hence the total ages of the most numerous WDs. The WD population most strongly constrains the IFMR zero-point at low masses, through the offset $b_0$ (\secref{sec:calibrated_model}), while the low-mass slope remains close to its prior. The inferred low-mass behaviour is broadly consistent with recent empirical determinations \citep{Cummings2018, ElBadry2018, Cunningham2024, Miller2026}, though residual degeneracies between the IFMR, IMF, SFH, and cooling models remain. Improved empirical IFMR constraints from wide binaries, star clusters, and volume-limited WD samples will be essential for reducing these systematics.

\subsection{Absence of binary evolution}

A limitation of the present model is the absence of binary evolution, low-mass helium-core WDs, unresolved WD+MS systems, double degenerates, and merger products are not modelled. This affects both the low-mass WD population, which we exclude below $\sim0.6\,M_\odot$ (\secref{sec:calibrated_model}), and the high-mass tail of the WD mass distribution, where mergers can contribute a significant fraction \citep{Althaus2013, Cunningham2024}. A substantial share of WD progenitors are members of binary systems \citep{Toonen2017, Torres2022, SantosGarcia2025}, so this omission is a genuine source of systematic uncertainty, particularly for the WD mass function; incorporating binary evolution is a priority for future versions of \textsc{CALOMERA}.

\subsection{Model-dependence of the Galactic predictions}

The predicted WD number densities and survey yields are forward-model forecasts and inherit the assumptions of the radially extended model, the radial SFH, AMR, IMF, extinction model, and cooling prescription. This model-dependence is strongest away from the Solar neighbourhood, where the present-day WD density reflects the integrated star formation at each Galactocentric radius, and where the radial SFH and AMR are adopted from \citet{Sysoliatina2022} rather than constrained by the present fit. The predictions therefore constitute a physically motivated baseline; different assumptions about inside-out growth or radial gradients would yield quantitatively different counts.

\subsection{Radial migration}
\label{sec:migration_caveat}
Radial migration is not included in the model, so the inferred SFH represents the age distribution of stars currently near the Sun rather than the true in-situ SFH at $R_\odot$. Chemo-dynamical and simulation-based studies show that migration can substantially reshape local age distributions by transporting old inner-disc stars into the Solar neighbourhood \citep{Minchev2026, Bernaldez2026, Ratcliffe2026}. We omit this correction deliberately, to allow direct comparison with previous WD-based SFH studies, most of which likewise do not correct for migration; incorporating it self-consistently is an important next step.

\subsection{Extension to the substellar regime}
The joint calibration provides a self-consistent description of the Galactic stellar population across the full main sequence and WD sequence. Because the underlying Galactic model is independent of the adopted stellar evolution library, the same framework can naturally be extended to other populations. As a first demonstration, we apply CALOMERA to the local brown-dwarf population using the calibrated IMF and SFH.

As brown dwarfs cool continuously and are sensitive to the low-mass IMF and the SFH, the same calibration predicts their local population without additional tuning. Our ATMO-based predictions indicate that current censuses may still miss a substantial fraction of the coolest T and Y dwarfs (\secref{sec:results_bd}), and that foreground brown dwarfs can contribute to compact red-source samples in deep \textit{JWST} fields \citep{Langeroodi2023, Hainline2024, Tu2025}. Because the substellar population was not part of the calibration, and because the predicted counts are sensitive to the low-mass IMF slope, these should be regarded as predictions of the Galactic model rather than as constraints on the substellar mass function.

\section{Conclusions}
\label{Sec:Conclusions}

We have presented \textsc{CALOMERA}, a Galactic population synthesis framework that combines the semi-analytic Milky Way disc model of \citetalias{Sysoliatina2021} with the stellar evolution from PARSEC isochrones combined with the three white dwarf cooling models: Montréal, BaSTI, and LPCODE. The framework self-consistently predicts the colour--magnitude distribution of both non-degenerate stars and white dwarfs, enabling the simultaneous modelling of the Galactic stellar populations, the star formation history, the initial mass function, the initial--final mass relation, and the white dwarf cooling physics. The parameters are constrained through an MCMC analysis of the observed \textit{Gaia} colour--magnitude diagram. Recalibration with the included WD population reduced the WD-count residual from 81\%, as predicted by \citetalias{Sysoliatina2021}, to sub-1\% while preserving the agreement with the main-sequence and giant populations, reproducing the WD cooling sequence, local luminosity function, and mass distribution simultaneously (see \secref{sec:results_sj21}).

Our main results of this recalibration are as follows:

\begin{itemize} 

\item The joint fit of SFH, IMF, and WD model parameters recovered a star formation history that peaks early and declines towards the present, while retaining the two recent star formation enhancement episodes recovered in \citetalias{Sysoliatina2021}. The calibration places the stronger of these episodes at approximately $\tau_\mathrm{p1}\approx2.4$\,Gyr, while the more recent one at \mbox{$\tau_\mathrm{p2}\approx0.5$\,Gyr} was adopted from \citetalias{Sysoliatina2021}. The overall declining SFH is consistent with recent CMD-based reconstructions of the extended Solar neighbourhood and contrasts with the near-constant or rising histories inferred from several white dwarf-only studies. Because the same SFH is required to simultaneously reproduce all populations including white dwarfs, our recovered star formation history is more tightly constrained than in single-population analyses, although it preserves sensitivity to the adopted stellar models, the assumptions on the Galactic structure, the likelihood construction, and other systematic uncertainties (see \secref{sec:SFR_IMF_IFMR} and \figref{fig:ifmr-imf-sfr}).

\item The new model favours an IFMR which in the low-mass range of initial masses predicts WD masses which are somewhat higher than those predicted by other IFMR from the literature. Thus, our IFMR indicates less intense mass loss at the stage of planetary nebula. Our data does not provide any additional constraints on the intermediate- and high-mass IFMR regimes. Our updated fraction of DA-type white dwarfs, 0.77, remains very close to its widely used value of 0.8. The simulation also does not indicate that the cooling process occurs slower than prescribed by the tested WD cooling models.  

\item The calibrated IMF remains close to the canonical Milky Way parameterisations, following a Kroupa-like slope above $\sim1\,M_\odot$, approaching the Chabrier form at the highest masses, and exhibiting a modest flattening below $1\,M_\odot$ (see \secref{sec:SFR_IMF_IFMR} and \figref{fig:ifmr-imf-sfr}).

\end{itemize} 

We also took several steps in modelling the CMD faint end across the Galaxy to provide observation-oriented predictions potentially useful for survey plannings: 

\begin{itemize} 
\item Using the calibrated model, we predicted the $R$--$z$ distribution of WDs throughout the disc and their observable fraction as a function of extinction, photometric band, and survey depth. For example, a survey reaching $G \simeq 26$ mag recovers about 95\% of WDs within 1~kpc, while a comparably complete census within 2~kpc requires $G \simeq 28$ mag. The SDSS $i$ band provides similar or slightly higher completeness, reaching about 97\% within 1~kpc at $i \simeq 26$ mag and 98\% within 2~kpc at $i \simeq 28$ mag (see \secref{sec:PredictionOfModel} and \tabref{tab:wd_completeness_allbands}).

\item Using the same forward-modelling framework, we obtained predictions for the Galactic brown-dwarf population based on ATMO models. The obtained 25-pc star counts agree within 20\% with the CNS5 data that is an encouraging result taking into account that brown dwarfs were not included into the calibration process. We also provided the expected numbers of L, T, and Y dwarfs in infrared surveys and their contribution as foreground contaminants to deep \textit{JWST} extragalactic observations, where such sources can be mistaken for compact high-redshift galaxies (see \secref{sec:results_bd}).
\end{itemize} 

Future developments will focus on improving the dynamical treatment of the Milky Way disc, providing three-dimensional kinematics, incorporating the radial migration effects, and adding distant binaries to the population synthesis module. Together with the next generation of deep astrometric and photometric surveys, such as, LSST \citep{Ivezic2019_LSST}, the Nancy Grace Roman Space Telescope \citep{Spergel2015}, GaiaNIR \citep{Hobbs2016}, and Euclid \citep{Laureijs2011}, and with increasingly complete catalogues of white dwarfs, lower main-sequence stars, and brown dwarfs, these improvements will strengthen the use of these faint low-mass objects as complementary tracers of the formation and evolution of the Milky Way.

\section*{Acknowledgements}

This work was funded by the Deutsche Forschungsgemeinschaft (DFG, German Research Foundation) through grant SY 230/1-1; part of this work was supported by Project-ID 138713538 – SFB 881 (`The Milky Way System'), subproject A06.
KS thanks the Institute for Astrophysics Potsdam (AIP) for providing computing resources for this work. 
AV would like to thank Andreas Just, Stefan Jordan, Maria Camisassa, Pier-Emmanuel Tremblay, Simon Blouin, and Maurizio Salaris for their discussions. A part of the analysis was carried out on the computing cluster at the Max Planck Institute for Astrophysics (MPA) and on systems at the Max Planck Computing and Data Facility (MPCDF). 
The authors used generative artificial intelligence tools (including Anthropic Claude, AstaAI, and Grammarly) to assist with grammar editing, code editing, and preliminary literature searches. All scientific analyses, interpretations, and conclusions were performed and verified by the authors.
The authors thank Oleksiy Golubov for providing useful comments on the manuscript. 
This work has made use of the following software and services: \texttt{Astropy} \citep{AstropyCollaboration2018}, \texttt{SciPy} \citep{Virtanen2020_SciPy}, \texttt{TOPCAT} \citep{Taylor2005_TOPCAT}, \texttt{Seaborn} and \texttt{Matplotlib} \citep{Hunter2007_Matplotlib, Waskom2021_Seaborn}, and the NASA Astrophysics Data System (ADS).

Last but not least, the authors thank the anonymous referee whose valuable comments helped to improve the quality of this paper. 

\section*{Data Availability}

The Jupyter notebooks containing the analyses presented in this work can be accessed at \href{https://github.com/askenja/gaia-wd-fitting}{https://github.com/askenja/gaia-wd-fitting}. The source code of the latest version of the JJ model, including the CALOMERA extension for white dwarfs and brown dwarfs, is hosted at \href{https://github.com/askenja/jjmodel}{https://github.com/askenja/jjmodel}. The WD isochrones generated for this study can also be accessed through the JJ model code. The simulation data underlying this work can be obtained from the corresponding authors upon request.



\bibliographystyle{mnras}
\bibliography{references_wd} 



\appendix

\section{Hess diagram comparison}
\label{app:HessComparison}
\begin{figure*}
    \centering
    \includegraphics[width=1\linewidth]{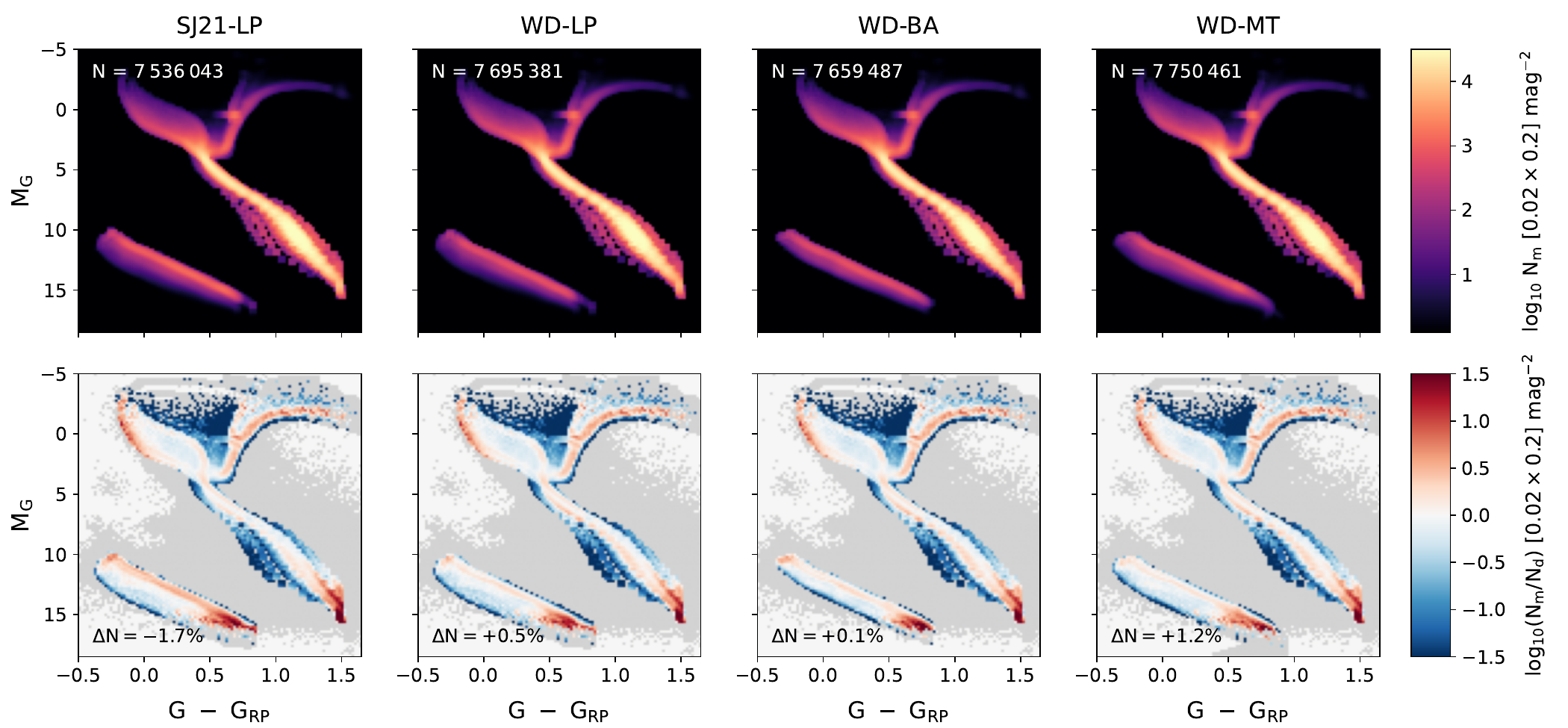}
    \caption{Comparison of the predicted Hess diagrams for the legacy model (SJ21-LP) and the three recalibrated models (WD-LP, WD-BA, and WD-MT). The upper row shows the predicted stellar density in the colour--magnitude diagram, and the lower row shows the logarithmic residuals with respect to the observed \textit{Gaia} sample. The recalibrated models substantially reduce the residuals along the white dwarf cooling sequence while preserving the agreement with the non-degenerate stellar populations.}
    \label{fig:hess-4models}
\end{figure*}

Figure~\ref{fig:hess-4models} compares the predicted Hess diagrams obtained with the legacy calibration of \citetalias{Sysoliatina2021} extended to WDs (SJ21-LP) and the three recalibrated models based on the LPCODE (WD-LP), BaSTI (WD-BA), and Montréal (WD-MT) white dwarf cooling sequences. The upper panels show the predicted stellar density in the colour--magnitude diagram, while the lower panels display the logarithmic residuals with respect to the observed \textit{Gaia} sample. Positive (red) and negative (blue) residuals indicate regions where the model over- or underpredicts the observed number density.

\section{Corner plot and additional MCMC results}
\begin{figure*}
    \centering
    \includegraphics[width=0.88\linewidth]{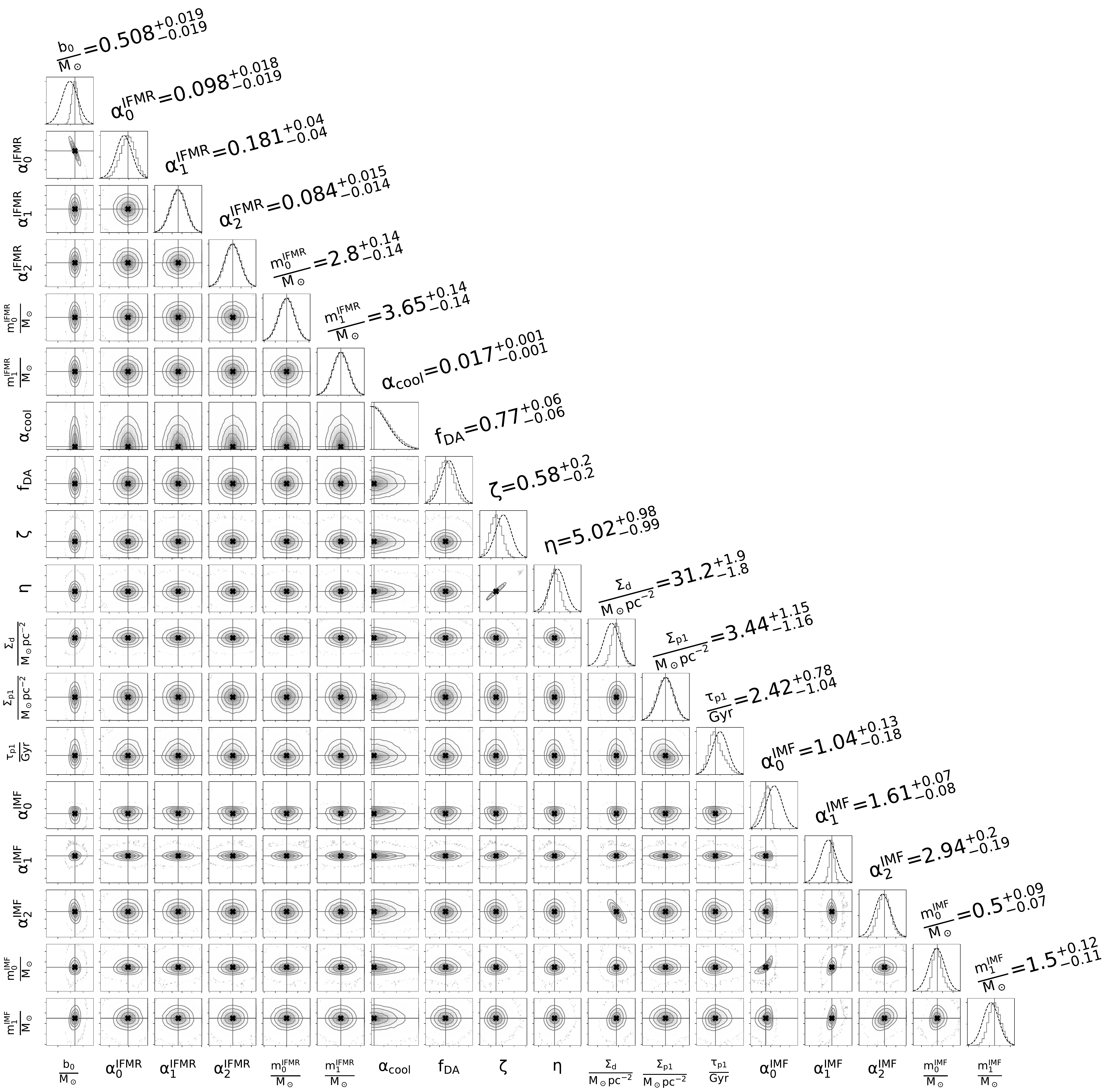}
    \caption{Full corner plot for the WD-LP calibration. Diagonal panels show the marginal posterior probability distributions of the fitted parameters, with median values and $1\sigma$ credible intervals. Off-diagonal panels show the corresponding two-dimensional posterior distributions and parameter correlations. The complete set of free parameters sampled during the MCMC optimisation is shown.}
    \label{fig:FullCornerPlot}
\end{figure*}

Figure~\ref{fig:FullCornerPlot} presents the full posterior distribution of the free parameters for the WD-LP calibration. The diagonal panels show the marginalised one-dimensional posterior distributions, with the quoted median values and corresponding 16th and 84th percentiles. The off-diagonal panels show the joint posterior distributions, illustrating the correlations and degeneracies between the fitted parameters.

\section{Brown dwarf predictions}
\label{app:BDPredictions}

\begin{figure*}
    \centering
    \includegraphics[width=0.88\linewidth]{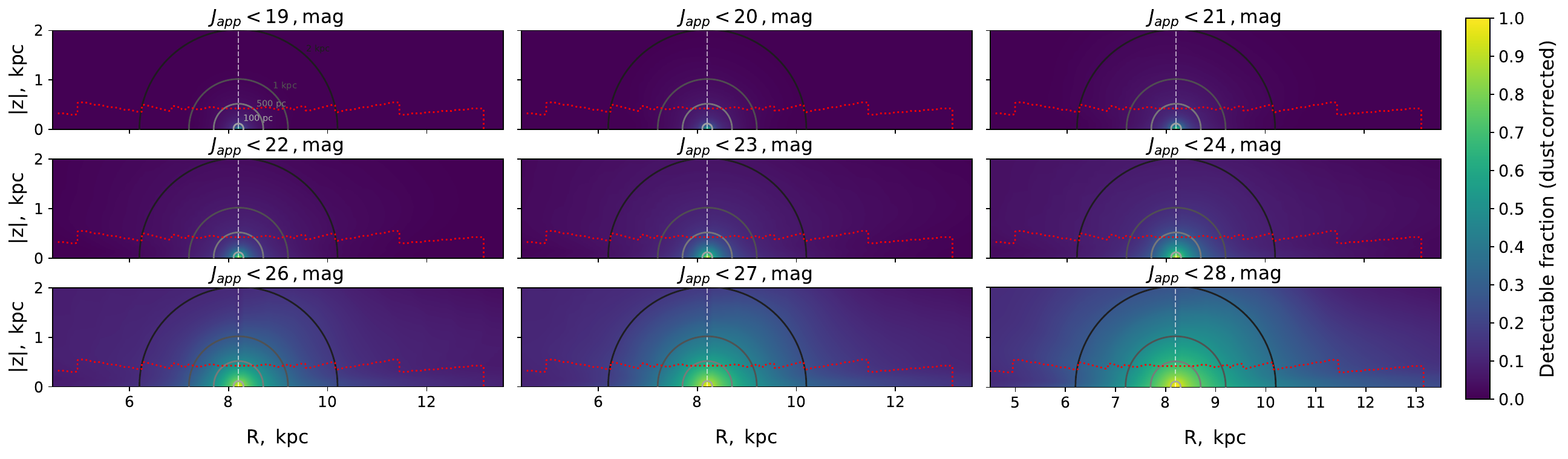}
    \caption{Predicted detectable fraction of the Galactic brown dwarf population in the MKO $J$ band as a function of Galactocentric radius and height above the Galactic plane, including interstellar extinction from the three-dimensional dust map of \citet{Lallement2022}. The panels show limiting apparent magnitudes from $J \in (19,28)$ mag, illustrating the increasing completeness achieved by deeper surveys.}
    \label{fig:BD_J_detection}
\end{figure*}

In addition to the white dwarf predictions presented in the main text, the calibrated model provides estimates of the detectability of the Galactic brown dwarf population in commonly used near- and mid-infrared passbands. Figure~\ref{fig:BD_J_detection} shows the predicted detectable fraction in the MKO $J$ band, including extinction from \citet{Lallement2022}. Owing to the intrinsically faint $J$-band luminosities of cool T and Y dwarfs, progressively deeper observations are required to achieve high completeness.

To illustrate a practical application of the model, Table~\ref{tab:JWST_BD_counts} presents the predicted cumulative foreground counts of L, T, and Y dwarfs in several representative \textit{JWST} extragalactic survey fields. The calculations use synthetic F200W photometry obtained from the ATMO CEQ NIRCam module~A models and are based on the fiducial WD-LP calibration. 
For each survey, the published field centre in Galactic coordinates and the survey solid angle are adopted to define the corresponding line of sight. The sightline is sampled in heliocentric distance shells, which are mapped onto the corresponding Galactocentric radius and height above the Galactic plane. The stellar density is evaluated using the nearest radial annulus of the Galactic model and integrated over the shell volume.
The predicted foreground population evolves systematically with survey depth: L dwarfs dominate the bright counts, while T dwarfs become the largest contributor at intermediate depths, and Y dwarfs appear in significant numbers only in the deepest observations. These predictions provide an estimate of the expected Galactic foreground contamination in ultra-deep \textit{JWST} imaging and complement recent observational studies identifying brown dwarfs among compact red sources.

\setlength{\extrarowheight}{0.25em}
\begin{table*}
\centering
\small
\caption{Predicted cumulative numbers of foreground L, T, and Y dwarfs in representative \textit{JWST} extragalactic survey fields as a function of limiting F200W magnitude. The predictions are based on the fiducial WD-LP model using synthetic F200W photometry from the ATMO models and do not include interstellar extinction. Note that individual spectral-type counts and totals are rounded independently to the nearest integer after summing the fractional model counts and may therefore differ slightly.}
\label{tab:JWST_BD_counts}
\begin{tabular}{crrrrrrrrrrrrr}
\toprule
&
\multicolumn{4}{c}{JADES} &
\multicolumn{4}{c}{CEERS} &
\multicolumn{4}{c}{COSMOS-Web} \\
\cmidrule(lr){2-5}\cmidrule(lr){6-9}\cmidrule(lr){10-13}
$F200W_{\rm lim}$ &
L & T & Y & Total &
L & T & Y & Total &
L & T & Y & Total \\
\midrule

22 &
3 & 0 & 0 & 3 &
2 & 0 & 0 & 2 &
40 & 4 & 0 & 45 \\

24 &
7 & 3 & 0 & 10 &
4 & 2 & 0 & 6 &
118 & 39 & 1 & 157 \\

26 &
9 & 13 & 1 & 22 &
5 & 7 & 0 & 12 &
160 & 195 & 8 & 363 \\

28 &
9 & 30 & 5 & 44 &
5 & 16 & 3 & 24 &
162 & 500 & 73 & 735 \\

30 &
9 & 38 & 21 & 68 &
5 & 21 & 12 & 38 &
162 & 713 & 325 & 1200 \\

32 &
9 & 38 & 42 & 89 &
5 & 21 & 24 & 49 &
162 & 724 & 713 & 1599 \\
\bottomrule
\end{tabular}
\end{table*}

\bsp	
\label{lastpage}
\end{document}